\documentclass[fleqn,usenatbib]{mnras}

\usepackage{amsmath}	
\usepackage{txfonts}
\usepackage[super]{nth}

\usepackage[T1]{fontenc}

\DeclareRobustCommand{\VAN}[3]{#2}
\let\VANthebibliography\thebibliography
\def\thebibliography{\DeclareRobustCommand{\VAN}[3]{##3}\VANthebibliography}

\usepackage{graphicx}	
\usepackage{amssymb}	
\usepackage[separate-uncertainty=true]{siunitx}

\title[Crosstalk in image plane beam combination]{Crosstalk in image plane beam combination for optical interferometers}

\author[D. J. Mortimer et al.]{
Daniel J. Mortimer,$^{1,2}$\thanks{E-mail: d.j.mortimer@exeter.ac.uk}
David F. Buscher$^{1}$
\\
$^{1}$Astrophysics Group, Cavendish Laboratory, J.J. Thomson Avenue, Cambridge CB3 0HE, UK\\
$^{2}$Department of Physics and Astronomy, University of Exeter, Physics Building, Stocker Road, Exeter EX4 4QL, UK
}

\date{Accepted XXX. Received YYY; in original form ZZZ}

\pubyear{2022}

\begin{document}
\label{firstpage}
\pagerange{\pageref{firstpage}--\pageref{lastpage}}
\maketitle

\begin{abstract}
Image plane beam combination in optical interferometers multiplexes the interference fringes from multiple baselines onto a single detector. The beams of starlight are arranged in a non-redundant pattern at the entrance of the combiner so that the signal from each baseline can be separated from one another in the frequency domain. If the signals from different baselines overlap in the frequency domain, this can give rise to a systematic error in the fringe measurements known as baseline crosstalk. In this paper we quantify crosstalk arising from the combination of atmospheric seeing and beam propagation over distances of order hundreds of metres. We find that in idealised conditions atmospheric wavefront errors and beam propagation do not contribute to crosstalk. However, when aperture stops are included in the optical beam train we observe that wavefront errors can result in squared visibility errors arising from crosstalk as high as $\Delta V^{2}$~=~\num{6.6e-3} under realistic observing conditions. 

\end{abstract}

\begin{keywords}
instrumentation: high angular resolution -- instrumentation: interferometers
 -- atmospheric effects -- methods: observational -- techniques: interferometric
\end{keywords}



\section{Introduction}

Long baseline interferometers such as the Very Large Telescope Interferometer (VLTI) \citep{2012SPIE.8445E..0DH}, Center for High Angular Resolution Astronomy (CHARA) \citep{2005ApJ...628..453T}, Navy Precision Optical Interferometer (NPOI) \citep{1998ApJ...496..550A}, and, currently under construction, the Magdalena Ridge Observatory Interferometer (MROI) \citep{2018SPIE10701E..06C} enable observations with milliarcsecond spatial resolution, the highest currently possible at optical wavelengths. This is achieved by transporting starlight from two or more interferometric collectors, typically meter class telescopes \citep[e.g.][]{2018SPIE10701E..0OB}, to a beam-combining laboratory where an instrument, referred to as a beam combiner, combines the light in some form to produce interference fringes. One form of beam combination is image plane beam combination which produces spatially-encoded interference fringes. From these fringes the interferometric observables, primarily visibilities and (in the case of the light from three or more telescope being combined) closure phases, can be extracted. 

Image plane combination has been implemented in many beam combiners \citep{2007A&A...464....1P, 2012SPIE.8445E..0MG, 2014Msngr.157....5L, 2020AJ....160..158A, 2020SPIE11446E..0VM} with applications of the technique dating as far back as the \nth{19} century when it was used by \cite{1891Natur..45..160M} to measure the angular diameter of Jupiter's moons. In such a scheme the telescope pupils are arranged on a non-redundant spacing pattern and are eventually brought to a focus by the beam combining instrument where the resulting Point Spread Function (PSF) is modulated by interference fringes, with one set of fringes for each pair of pupils. The non-redundant spacing ensures that the interference fringes from each pupil pair are sampled at sufficiently different spatial frequencies such that the visibilities resulting from each combination pair can be measured independently.

When multiple baselines are measured simultaneously in a single combiner there is a potential for a systematic error known as crosstalk. This is where information from one baseline mixes with that of another, altering the measured visibility for a given baseline, which can lead to incorrect interpretations of the data.

The level of crosstalk which can be tolerated depends on the scientific application. An example application is the study of the photosphere of red supergiant stars \citep{2009A&A...508..923H, 2010A&A...515A..12C, 2017A&A...605A.108M, 2018A&A...614A..12M} where to study structure on the stellar disk, observations must be made beyond the first lobe of the visibility function, giving measured squared visibilities often as low as $V^{2}$~=~10$^{-2}$ with \cite{2017A&A...605A.108M} reporting measurements in the \nth{16} lobe of the visibility function and $V^{2}$ as low as $V^{2}$~=~10$^{-5}$. In this regime, very low levels of crosstalk of signals from baselines of high visibility ($V^2\sim 1$) to the low-visibility baselines can be significant.

In this paper we expand on the existing models of image plane beam combination crosstalk and quantify the levels of crosstalk arising when the effects of atmospheric seeing, diffraction due to free space propagation, finite sized optics and unequal path lengths are considered. In Section~\ref{Methods} we discuss both the conceptual design and practical implementation of the model used in this simulation, including parameter definitions and explored parameter space. In Section~\ref{Results} we discuss the results of our simulation, highlighting the effects of each parameter on the observed crosstalk in turn, and how to minimise crosstalk within an optical interferometer. In Section~\ref{Discussion} we apply our findings to two practical examples of observations with a three and five telescope beam combiner. Finally, in Section~\ref{Conclusions} we summarise our findings. 

\section{Methods} \label{Methods}

\subsection{Model} \label{Model}

The model used in our simulation is that of a simple two-telescope interferometer observing an unresolved point source, a source whose angular diameter is significantly less than $\lambda/B$ where $\lambda$ is the wavelength of observation and $B$ the baseline, or projected on-sky separation of the two telescopes, and hence our simulated observations have a ideal visibility of unity. While only two telescopes are simulated here, the results from this model can be used to infer the crosstalk in an interferometer with any number of telescopes using the fact that the propagation and interference of starlight on any given baseline is independent of the other baselines. 

\begin{figure}
	\includegraphics[width=\columnwidth]{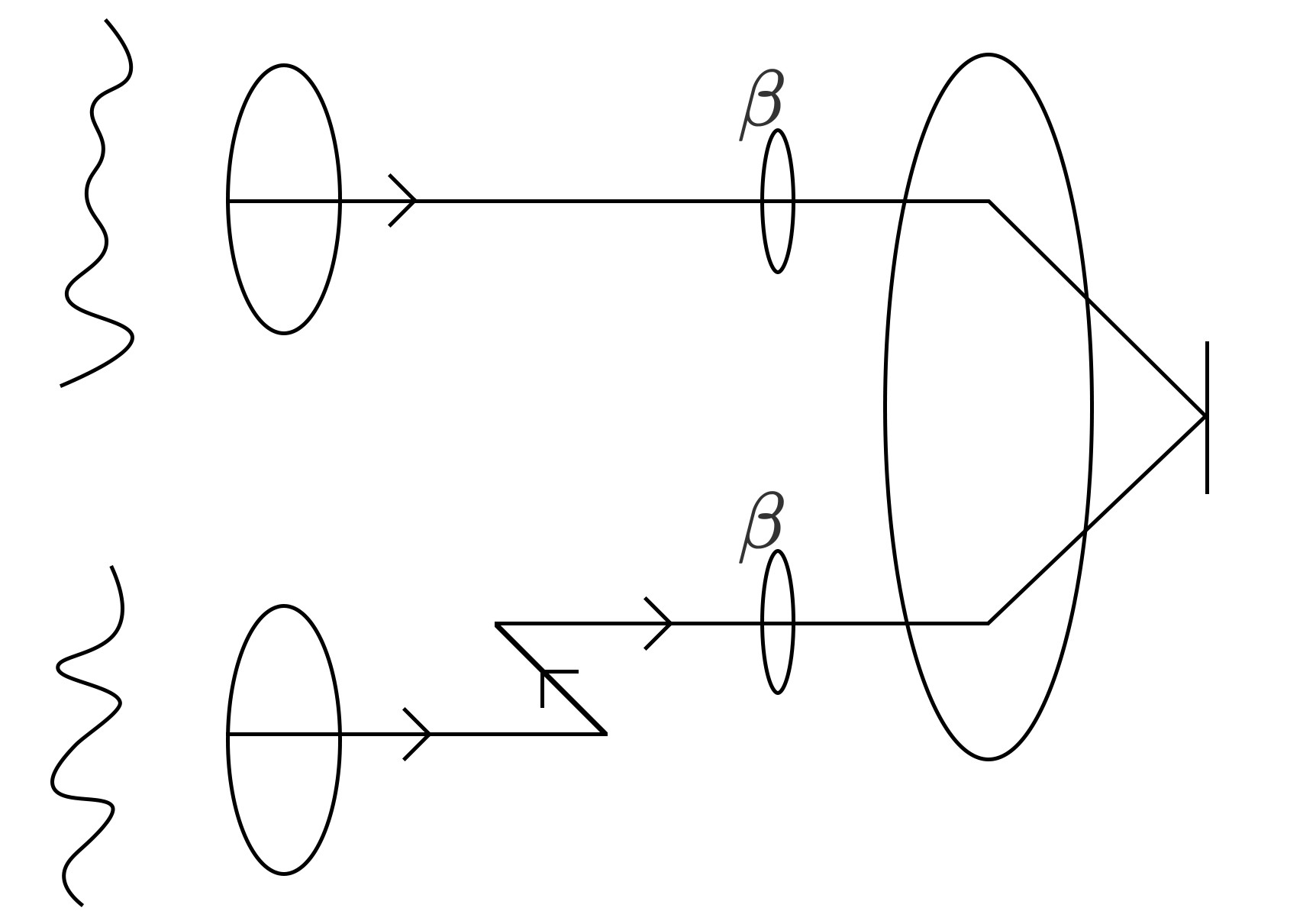}
    \caption{A simplified schematic demonstrating the various stages of the simulation. From left to right the atmospherically perturbed complex amplitudes are cropped by the circular collecting apertures before being optionally tip-tilt corrected and propagated, with the lower beam being propagated a greater distance. The two beams are then cropped by a circular aperture after propagation before being brought to a focus to generate a PSF modulated by interference fringes.}
    \label{schematic_of_simulation}
\end{figure}

The modelled interferometer is subject to atmospheric seeing. It is assumed that the telescopes are in the near field of the atmospheric turbulence and hence there are only phase perturbations and not amplitude perturbations in the received wavefront. The validity of this approximation is discussed by \cite{1974ApJ...189..587Y} and \cite{1981PrOpt..19..281R}. In the simulation the two phase perturbed wavefronts are collected by circular apertures and are optionally corrected for tip-tilt errors as all current long-baseline interferometers are equipped with at least tip-tilt level adaptive optics. The two perturbed wavefronts are then propagated to simulate the transporting of the wavefronts from the telescopes to the beam combining instrument. The two wavefronts can be transported by different distances to account for the unequal path lengths required in an interferometer for targets not at zenith. The beams are subsequently cropped by a scalable aperture, to simulate finite sized optics along the beam train, before they are brought to a focus to generate a PSF modulated by interference fringes which is the raw output of image plane beam combination. Fig.~\ref{schematic_of_simulation} outlines a simplified schematic of the simulation.

\begin{figure*}
	\includegraphics[width=\textwidth]{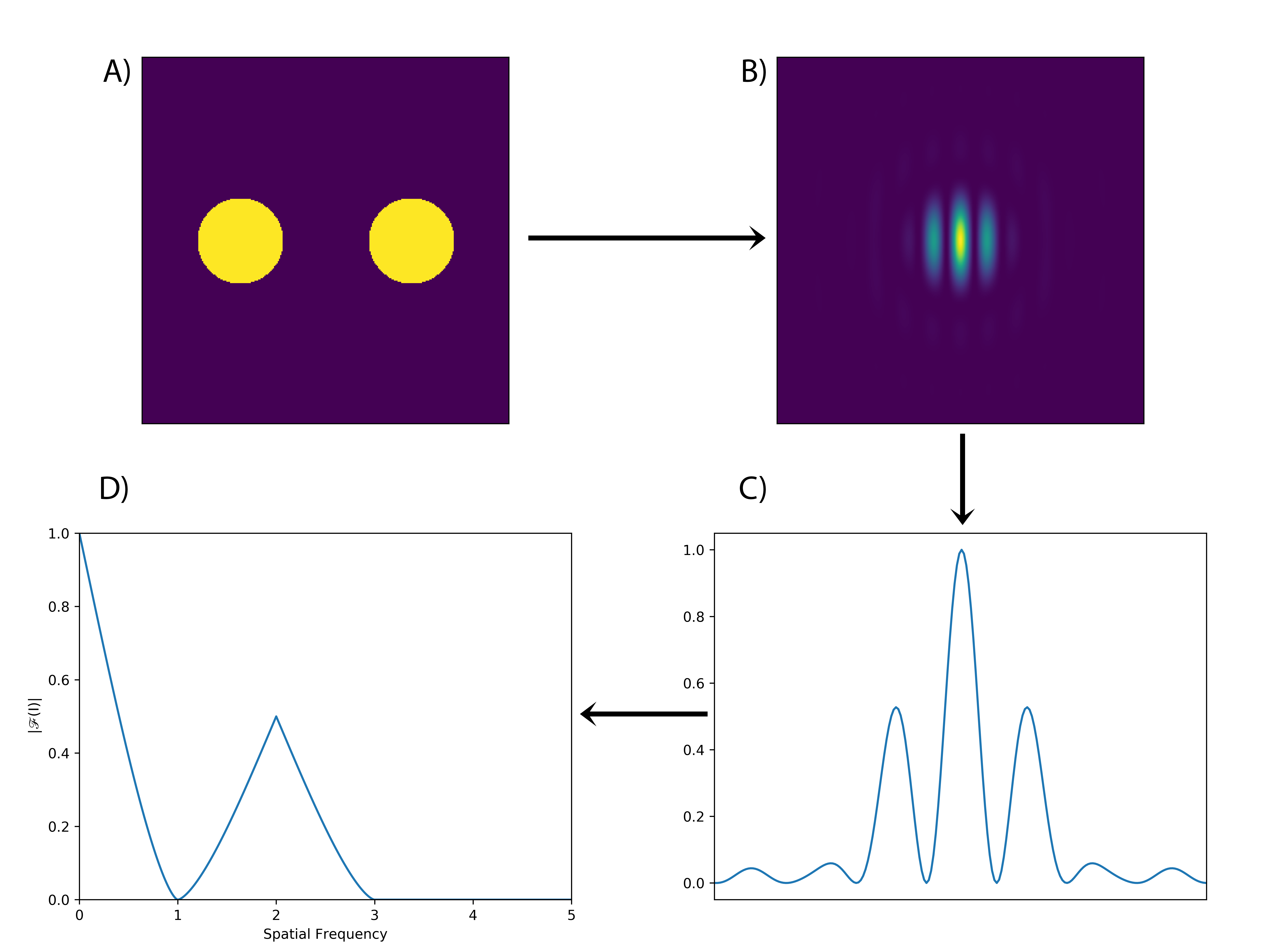}
    \caption{The process by which visibilities are extracted from the simulation for two pupils with no atmospheric perturbation or propagation, A) the two pupils, B) the PSF of the two pupils, generated by taking the squared modulus of the Fourier transformed pupil, C) the 1D LSF, generated by taking the average of each column of pixels (in the direction perpendicular to the fringes), D) the modulus of the Fourier transform of the LSF which is equivalent to the MTF, normalised here to the zero spatial frequency term. From the LSF the visibility can be extracted.}
    \label{pupil_plane_to_MTF}
\end{figure*}

From the PSF modulated by interference fringes the visibilities are then extracted. This process is designed to mimic how visibilities could be extracted in practice. We first average the PSF along a line parallel to the fringe peaks to yield a Line Spread Function (LSF) and then perform a 1D Fourier transform, taking the modulus of which results in the Modulation Transfer Function (MTF). The MTF contains two terms, the zero spatial frequency or DC term and the interference term originating from the interference of the two apertures. This slice of the MTF can equally well be obtained by taking the 2D Fourier transform of the PSF and extracting the 1D slice from the resulting 2D MTF along a line containing the DC term and the peak of the interference term. We assume that in the multi-baseline case the apertures are arranged in a linear fashion so that this slice contains the peaks of all the interference terms for all baselines. Fig.~\ref{pupil_plane_to_MTF} shows the output at the various stages of the simulation.

To understand the scale of the spatial frequency axis used in Fig.~\ref{pupil_plane_to_MTF} and throughout the rest of this paper we must first consider the relationship between the pupil function and the MTF. As discussed above, the MTF is derived from the pupil function via the PSF. The PSF is the square absolute of the Fourier transformed pupil function, $ b(f_x,f_y) \propto |\mathcal{F}\{P(x,y)\}|^2$ where $b(f_x,f_y)$ is the intensity pattern seen in the image plane and $P(x,y)$ the pupil function. The MTF is then the modulus of the Fourier transform of the PSF, i.e.\ MTF~$\propto |\mathcal{F}\{b(f_x,f_y)\}|$. Putting these together and using the Wiener–Khinchin theorem \citep{2000stop.book.....G} we get
\begin{equation}
    \text{MTF} \propto |\mathcal{F}^{-1}\{|\mathcal{F}\{P(x,y)\}|^2\}|\propto
    |\text{autocorr}(P(x,y))|.
\end{equation}
Hence the MTF derived from the process described above is the modulus of the normalised autocorrelation of the pupil function \citep{1995ApOpt..34.6337H}. 

\begin{figure*}
	\includegraphics[width=\textwidth]{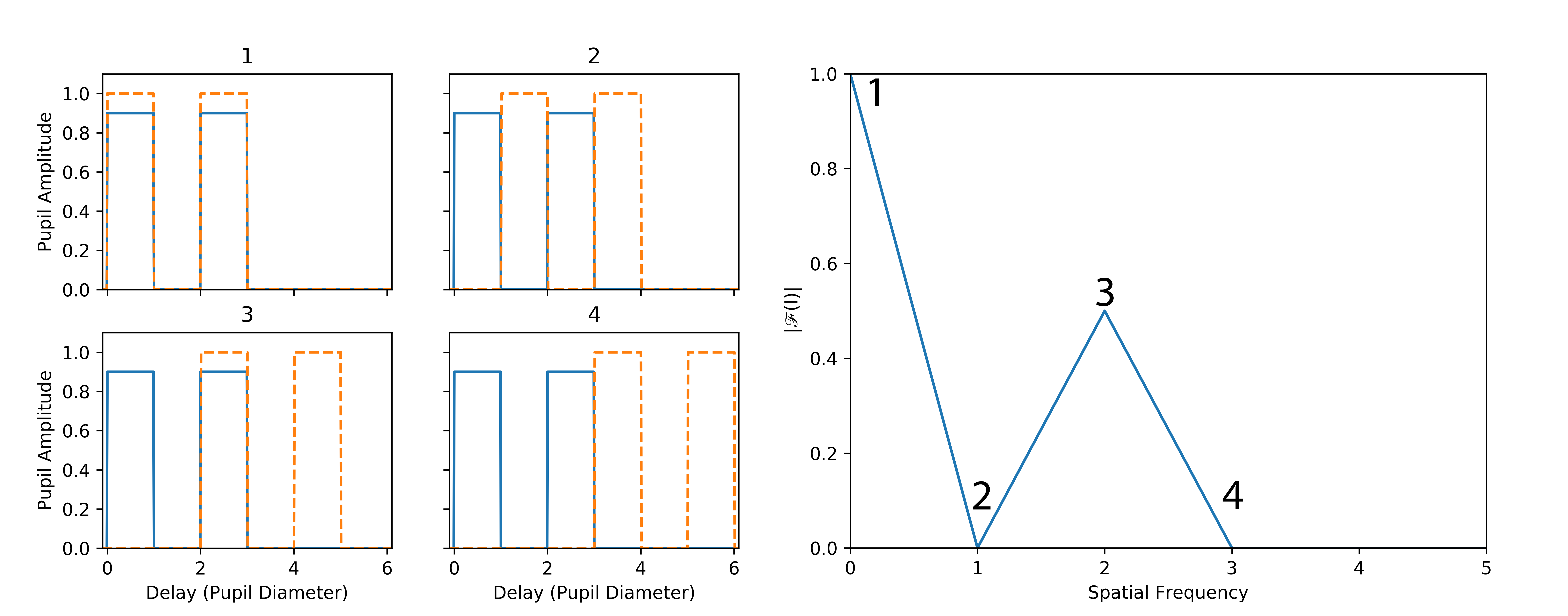}
    \caption{The autocorrelation of two rectangular functions separated by two pupil diameters with the pupil plane shown left and the resulting autocorrelation right. Four stages are labelled 1, the delay at zero giving a maximum value in the autocorrelation 2, with the delay at one pupil diameter 3, a delay of two pupil diameters 4, a delay of three pupil diameters. As the autocorrelation of the pupil function and MTF calculated in this paper are mathematically the same this then explains the definition of the spatial frequency axis used throughout the rest of this paper. Note the pupils in blue here have been reduced to an amplitude of 0.9 for the sake of clarity however, the pupils used throughout in this paper are of equal amplitude.}
    \label{Auto_corr_sliding_mtf}
\end{figure*}

We scale the spatial frequency axis of the MTF in terms of the diameter of the input apertures. The scaling can be understood by considering the MTF of a 1D version of two circular pupils, i.e.\ two 1D top hat functions. If we take two pupils spaced two beam diameters apart and perform the autocorrelation, as is done in Fig.~\ref{Auto_corr_sliding_mtf}, initially the autocorrelation is at a maximum value (normalised in the MTF to unity). This is the stage labelled 1 and is defined as zero spatial frequency. The spatial frequency of one is then defined when the autocorrelation has advanced one pupil diameter giving zero overlap and hence a MTF value of zero (stage 2). If the autocorrelation advances to two pupil diameters at what is defined as a spatial frequency of 2, the first pupil overlaps with the second and the MTF has half the magnitude as the zero spatial frequency component (stage 3). By the time the autocorrelation has swept through three pupil diameters there is no longer any overlap and the MTF again returns to zero (stage 4). This then is the explanation for the shape of the MTF and how the spatial frequency units relate to the pupil plane.

The effects of a dedicated spatial filter, which are typically implemented to remove wavefront errors and boost the observed visibility \citep{2001MNRAS.326.1381K}, are not considered unless explicitly mentioned. The effects of spatial filtering will be explored in-depth in a future paper (Mortimer et al., in prep). 

This simplified model does not account for the effects of a finite exposure time or optical aberration induced by the optics along a typical beam train. Ignoring these effects allows us to build a model which is applicable to any interferometer. 

\subsection{Classical crosstalk} \label{Classical_crosstalk}

\begin{figure}
	\includegraphics[width=\columnwidth]{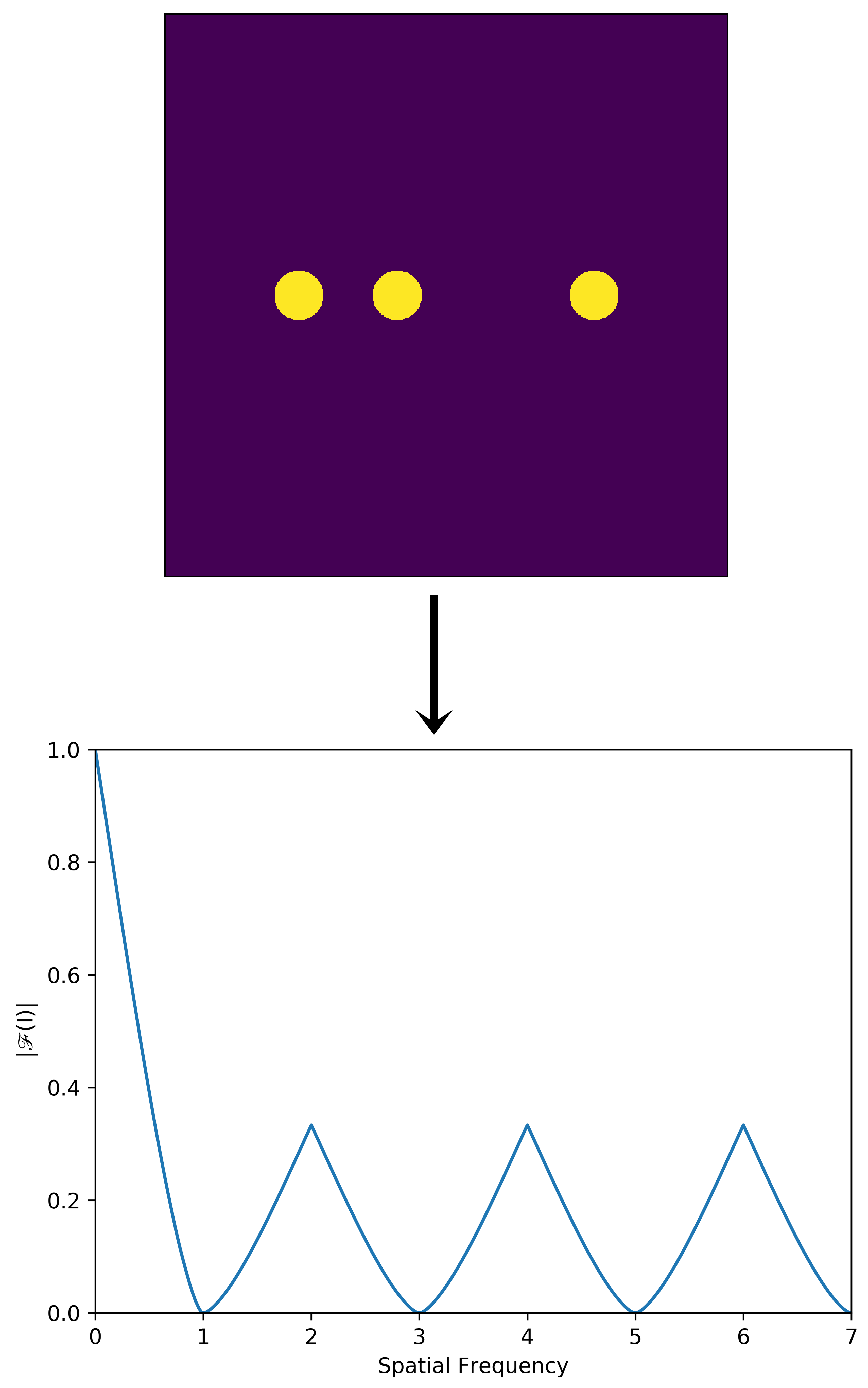}
    \caption{Top: three pupils well separated by 2$\times$, 4$\times$ and 6$\times$ the pupil diameter. Bottom: the resulting MTF derived from the PSF of the pupil plane above showing the DC term (0 < Spatial Frequency < 1) and the three interference terms corresponding to the three baselines with peaks at spatial frequencies of 2, 4 and 6.}
    \label{fft_mod_ideal_PSF_three_beams_2_4_6_seperation_combined}
\end{figure}

Here we discuss what has previously been considered when exploring crosstalk in the literature, what we call classical crosstalk, and how the work presented in this paper builds on this. In classical crosstalk the effects of atmospheric seeing and pupil propagation are not considered. However, this still gives rise to an interference term in the MTF with a significant width in spatial frequency space for each pupil combination pair or baseline (see Section~\ref{Model}) and therefore the spatial frequency of the other baselines must be sampled at sufficiently different spatial frequencies to ensure the information from different baselines are well separated. One implementation of sufficient spacing is shown in Fig.~\ref{fft_mod_ideal_PSF_three_beams_2_4_6_seperation_combined}. In this `2,4,6' arrangement the three pupils are separated from each other by 2$\times$, 4$\times$ and 6$\times$ the diameter of the undiffracted pupils. With three apertures there are three baselines and hence three interference peaks. In this `2,4,6' arrangement the peaks of the interference terms are well separated (with a null between them) and so would not experience crosstalk in the classical model. 

\begin{figure}
	\includegraphics[width=\columnwidth]{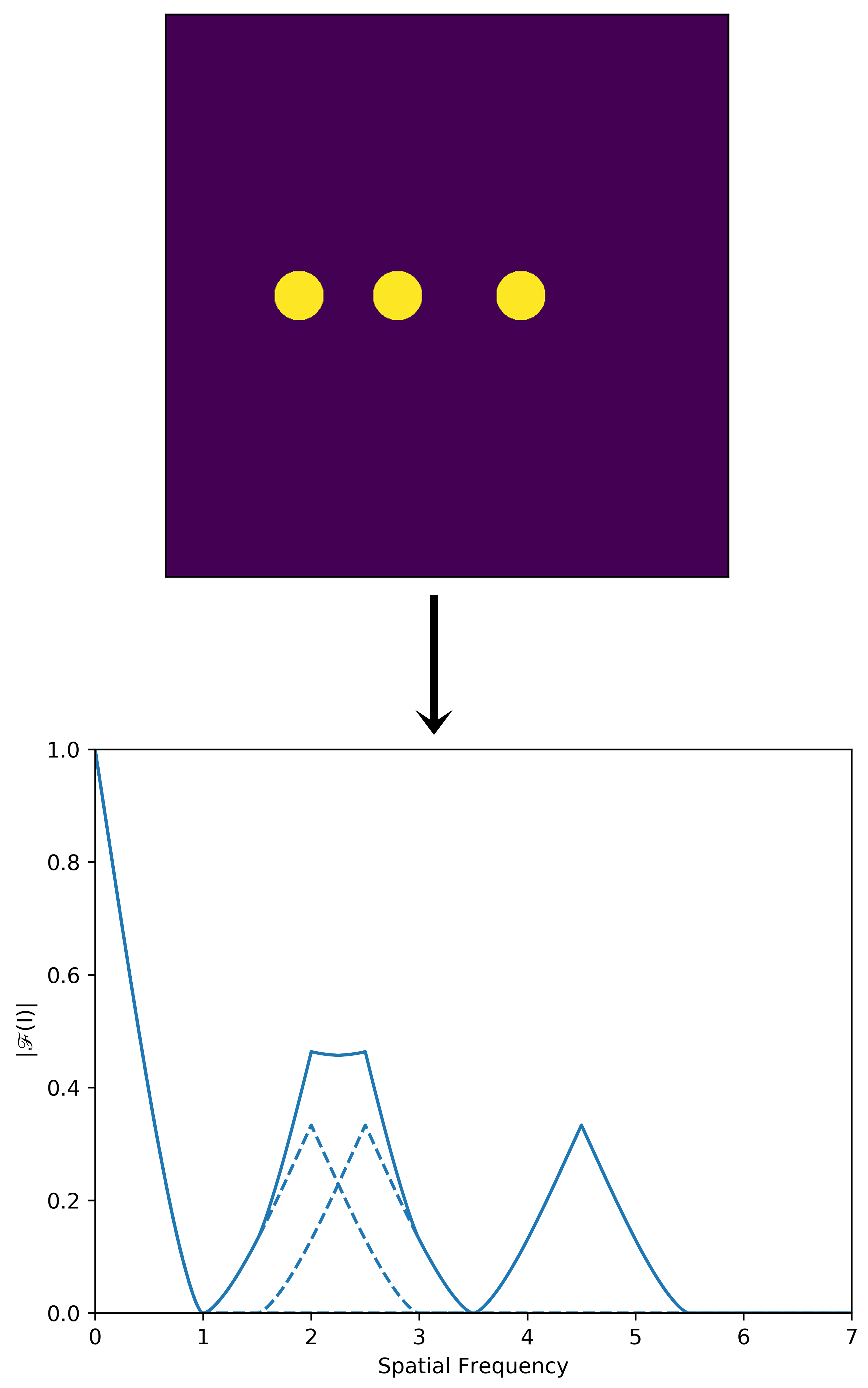}
    \caption{Top: three pupils separated by 2$\times$, 2.5$\times$ and 4.5$\times$ the pupil diameter. Bottom: the resulting MTF derived from the PSF showing the DC term (0 < Spatial Frequency < 1) and the three interference peaks corresponding to the three baselines peaking at spatial frequencies of 2, 2.5, 4.5. The profile of the individual interference terms are the same as Fig.~\ref{fft_mod_ideal_PSF_three_beams_2_4_6_seperation_combined} (represented by the dashed lines) however in the overall MTF (solid line) the lower two spatial frequency terms now overlap, boosting the observed peaks of the interference terms, demonstrating crosstalk.}
    \label{fft_mod_ideal_PSF_three_beams_2_2_point_5_4_point_5_seperation_combined}
\end{figure}

As an example of how crosstalk can arise in this classical model, if the right most pupil in Fig.~\ref{fft_mod_ideal_PSF_three_beams_2_4_6_seperation_combined} is moved 1.5 times the pupil diameter to the left such that the spacing between the pupils is now 2$\times$, 2.5$\times$ and 4.5$\times$ the pupil diameter as shown in Fig.~\ref{fft_mod_ideal_PSF_three_beams_2_2_point_5_4_point_5_seperation_combined} we observe a significant level of crosstalk. The interference terms of the two lower spatial frequency baselines (represented by the dashed lines) now overlap, resulting in the amplitude of the peaks at spatial frequencies of 2 and 2.5 appearing to be significantly greater in the overall MTF curve (represented by the solid line) than their true values, due to baseline crosstalk. 

This example also shows that in general crosstalk, represented by the symbol $\Delta V^{2}$ in this paper, is an additive error in $V^2$. Where power from baselines other than the one being measured in the MTF contribute to the total power measured at a given spatial frequency. 

Alongside visibility, a core observable for long baseline optical interferometers is the closure phase, allowing some phase information of the target under observation to be recovered through the turbulent atmosphere. This paper focuses on the amplitude information in the MTF which, as we have demonstrated above, can cause crosstalk in visibility measurements. This gives us reason to believe that phase information could also mix between baselines, which may lead to biases in closure phase measurements. A detailed analysis of such biases is beyond the scope of this paper and will be addressed in a future work (Mortimer et al., in prep). Instead in this paper we will focus only on the crosstalk effects on visibility measurements.

\subsection{Parameter space} \label{Parameter_space}

In order to make our results readily applicable to any interferometer our simulation is built using five dimensionless parameters. The range of values of each parameter in our simulation is guided by the expected values for that parameter at the MROI. The MROI will cover a wide range of parameter space, observing at wavelengths between \SI{0.6}{\micro\metre} and \SI{2.4}{\micro\metre}, on baselines from \SI{7.8}{\metre} to \SI{347}{\metre}. Most current-generation interferometers will also be covered by the parameter range used here. Four of the parameters are the same as those used in \citep{2001MNRAS.327..217H}, and are defined in the following paragraphs.

The parameter $\alpha$, given by
\begin{equation}
    \alpha = \dfrac{D}{r_0}
\end{equation}
characterises the telescope diameter ($D$) in terms of the Fried parameter ($r_0$): for a fixed-sized telescope it is inversely proportional to the strength of the seeing.
The values used here range from 1.5 to 16.5. Given the \SI{1.4}{\metre} aperture telescopes the MROI utilises, the worst case scenario for $\alpha$ (the largest value) was found by assuming a $r_0$ of $10.3$ cm (the equivalent of 1 arc-second of seeing) for a wavelength of $\lambda$~=~\SI{0.5}{\micro\metre} at zenith and then scaling $r_0$ to the shortest wavelength, $\lambda$~=~\SI{0.6}{\micro\metre}, and largest zenith angle (60$^{\circ}$) expected at the MROI. The best case scenario (smallest $\alpha$) was found by assuming a $r_0$ of 14.7 cm (the equivalent of 0.7 arc-seconds of seeing) for a wavelength of $\lambda$~=~\SI{0.5}{\micro\metre} at zenith, then scaling $r_0$ to the longest wavelength $\lambda$~=~\SI{2.4}{\micro\metre} and observations at the zenith. Converting $r_0$ between wavelengths from the seeing at $\lambda$~=~\SI{0.5}{\micro\metre} was calculated by the relation $r_0(\lambda) \propto (\lambda)^{6/5}$. Converting $r_0$ for a given zenith angle was calculated by $r_0(\theta)~=~r_0(\theta=0)(\cos\theta)^{3/5}$ \citep{1961wptm.book.....T}.

The parameter $\beta$, given by 
\begin{equation}
    \beta = \dfrac{d'}{d}.
\end{equation}
characterises the diameter of the cropping aperture after propagation $d'$ in terms of the diameter of the pupil before propagation $d$. The range of values for $\beta$ was chosen based on realistic ratios of beam size to the smallest clear aperture along a beam train, ranging from a $\beta$ of 1, where the optics have a clear aperture equal to the diameter of the collimated pupil before propagation, to a value of 3.

The parameter $\gamma$, given by
\begin{equation} \label{e:gamma}
    \gamma = \dfrac{z_1}{z_2}
\end{equation}
is the ratio of the propagation distance of one beam ($z_1$) to the other ($z_2$) and allows for the effects of unequal path lengths to be simulated. Here a value of $\gamma$ larger than unity is defined as $z_2$ propagating a distance shorter by a factor of $\gamma$ than $z_1$. At the MROI $\gamma$ ranges from 1 for observations at the zenith when the path length through both arms of the interferometer are equal, to 1.3 for observations when an object is observed at a zenith angle of 60$^{\circ}$ by two telescopes separated by a \SI{347}{\metre} baseline as part of a Y shaped array configuration.

The fourth dimensionless parameter $\delta$, given by 
\begin{equation} \label{delta_eq}
    \delta = \dfrac{4\lambda z_1}{\pi d^2}
\end{equation}
is a measure of how much the propagated beam is affected by diffraction, and includes the effects of wavelength $\lambda$, propagation distance $z_1$ and beam diameter $d$ in one parameter. Looking at equation~\eqref{delta_eq} it can be seen that the smallest $\delta$ arises for the shortest path length and the shortest wavelength. At the MROI this is for a baseline of \SI{7.8}{\metre} and a wavelength of \SI{0.6}{\micro\metre} which gives a $\delta$~=~0.07. Interestingly, for the MROI the contribution to $\delta$ due to the propagation from the telescope, along the beam relay system and through the delay lines is negligible in the case of the shortest baseline ($\delta$~=~0.006) compared to the contribution of the \SI{14.19}{\metre} propagation after the delay lines within the beam combination laboratory ($\delta$~=~0.064). This is because pupils travelling along the beam relay system and through the delay lines have a collimated diameter of \SI{95}{\milli\metre} whereas for propagation within the beam combination laboratory the pupils are compressed to a diameter of \SI{13}{\milli\metre}, see \cite{2013JAI.....240001B} for a further discussion of the MROI beam architecture. The largest value of $\delta$ at the MROI occurs at the wavelength of \SI{2.4}{\micro\metre} on the longest baseline (\SI{347}{\metre}), with the same \SI{14.19}{\metre} propagation distance after beam compression within the beam combination laboratory and is $\delta$~=~0.446.

The final dimensionless parameter is T, which defines the level of tip-tilt correction and ranges between 0, no correction, and 1, complete removal of tip-tilt atmospheric perturbations. 

\subsection{Propagation calculation} \label{Propgation_calculation}

The initial complex amplitudes are propagated by applying the Fresnel approximation of scalar diffraction theory as given in \cite{2005ifo..book.....G}. This is applicable to our simulation as the distances propagated (typically tens to hundreds of meters) are many times larger than the diameter of the beams being propagated (for the specific case of the MROI, beam diameters are \SI{95}{\milli\metre} propagated over tens to hundreds of meters), as well as the fact that the beam diameters are always significantly larger than the wavelength of light being propagated. The propagation is calculated in Fourier space by equation~\eqref{propagation_equation}.

\begin{equation} \label{propagation_equation}
    a(\hat{f_{x}}, \hat{f_{y}},z) = a(\hat{f_{x}}, \hat{f_{y}},0) \exp((-i\pi \delta)(\hat{f_{x}}^{2} + \hat{f_{y}}^{2})).
\end{equation}

This follows the form of \cite{2001MNRAS.327..217H} where $a(\hat{f_{x}}, \hat{f_{y}},0)$ represents the Fourier transform of the initial complex amplitude which is to be propagated a perpendicular distance z, $\delta$ is as defined in equation~\eqref{delta_eq} except that $d$ is the width of the grid being propagated, which is many times the diameter of the actual pupil to avoid aliasing, $\hat{f_{x}}$ and $\hat{f_{y}}$ are dimensionless spatial frequencies of the dimensionless spatial coordinates $\hat{x}~=~2x/d$ and $\hat{y}~=~2y/d$. The propagated beam is then recovered by taking the inverse Fourier transform of $a(\hat{f_{x}}, \hat{f_{y}},z)$.

\subsection{Visibility estimator} \label{Visibility_estimator}

Visibility is one of the key observables of an optical interferometer. In the case of image plane beam combination, the visibilities are extracted from the image formed by placing the beams from the telescopes side by side and focusing the light onto a detector. The image plane then contains an Airy disk modulated by interference fringes. By analysing the power contained in the interference term with respect to the zero spatial frequency term in the MTF of the PSF it is possible to extract the visibility. In the simulation the rms visibility for an ensemble of exposures was calculated using
\begin{equation} \label{Vis_equation}
    V^2_{\textrm{rms}} = 4\dfrac{\langle X^2\rangle}{\langle F\rangle^2},
\end{equation}
where $X$ represents the value of the unnormalised MTF at the expected central spatial frequency of the interference term, and $F$ the value of the unnormalised MTF at zero spatial frequency (the DC term), with $\langle \rangle$ denoting averaging over an ensemble of exposures. Note that, as used here, the MTF is not normalised to unity at the origin on an exposure-by-exposure basis, as for runs with a finite value of $\beta$ the value at the origin (the total flux) varies from exposure to exposure. This is due to the atmospheric phase perturbations causing amplitude variations as the beams are propagated, leading to a loss in flux as the pupils are subsequently truncated with a finite $\beta$.   For each run 10,000 unique realisations of an atmospherically perturbed observation were generated. These were subsequently distributed into 20 bins of 500 PSFs each. The value of $V^2_{\textrm{rms}}$ was then calculated for each of these bins resulting in 20 independent visibility estimates which were then averaged allowing for a final estimate of the visibility and its standard deviation to be generated for each run. From the standard deviation the standard error of the mean was calculated however it is often smaller than the line thickness of the plots in this paper. Due to how this technique was implemented it was possible to calculate a value for the visibility for all spatial frequencies simultaneously, and not just the visibility at the spatial frequency of the peaks in the MTF.

\subsection{Model implementation} \label{Model_implmentation}

Our atmospheric model is built using the \textsc{python} package \textsc{MegaScreen} \citep{2016OExpr..2423566B} which generates a single layer of phase perturbations (a phase screen) with a Von K\'{a}rm\'{a}n spectrum, characterised by the Fried parameter $r_{0}$ and an outer scale $L_0$. For all the following calculations $L_0$ was set to 58 times the aperture diameter for the majority of cases where the pupil was 120 pixels.

\textsc{MegaScreen} generates one or more `windows' on a infinite virtual phase screen where for each iteration of the phase screen generator one or more 2D arrays of phase screen perturbations are returned. Between iterations the frozen-turbulence model is advanced by a configurable step size to simulate time evolution. 

In our implementation two square windows of width equal to the diameter of the telescope (which was set to 120 pixels unless otherwise stated in the results presented below) were simulated per iteration, with the step size between iterations being set to the diameter of the telescope to ensure that each iteration was independent (i.e.\ did not share the same phase screen as the previous iteration). For the purposes of phase screen generation the two telescopes in our simulation were separated by 160 times the pupil diameter to ensure that the phase screens generated for the two apertures were also independent. 

The phase screens were then corrected in some runs for tip-tilt perturbations by the use of a Zernike filter. A Zernike filter is an approximation for an adaptive optics system in which various Zernike polynomials (in this case tip and tilt) are perfectly removed from the phase screen. This essentially simulates a perfect adaptive optics corrections up to the highest order Zernike removed and no adaptive optics corrections for higher Zernike orders.  

The pupils were then compressed from the diameter of the telescope ($D$) to a diameter $d$ (this is achieved in practice by redefining the pixel scale of the complex amplitudes, assuming perfect beam compression) and propagated through free space at the centre of a 3,000 pixel grid a distance $z_1$ and $z_2$ for the two beams respectively by the method outlined in Section~\ref{Propgation_calculation}. After propagation, the pupils were cropped by a circular aperture to simulate the effects of finite-sized optics along the beam train. 

Finally, the pupils were placed onto a common grid and converted to a PSF via a 2D Fast Fourier Transform (FFT) routine to simulate transformation to the image plane and beam combination. The resulting PSF, modulated by interference fringes, was then averaged along the direction perpendicular to the fringes to generate a 1D LSF. Taking the absolute value of the FFT of the LSF gives the MTF. The value of the MTF at a spatial frequency of zero was stored for each iteration of the simulation and used as the $F$ term in equation~\eqref{Vis_equation}. Equation~\eqref{Vis_equation} was then computed for each spatial frequency sampled in the MTF by the procedure discussed in Section~\ref{Visibility_estimator} with $X$ being the amplitude of the MTF at the spatial frequency being calculated.

\begin{figure}
	\includegraphics[width=\columnwidth]{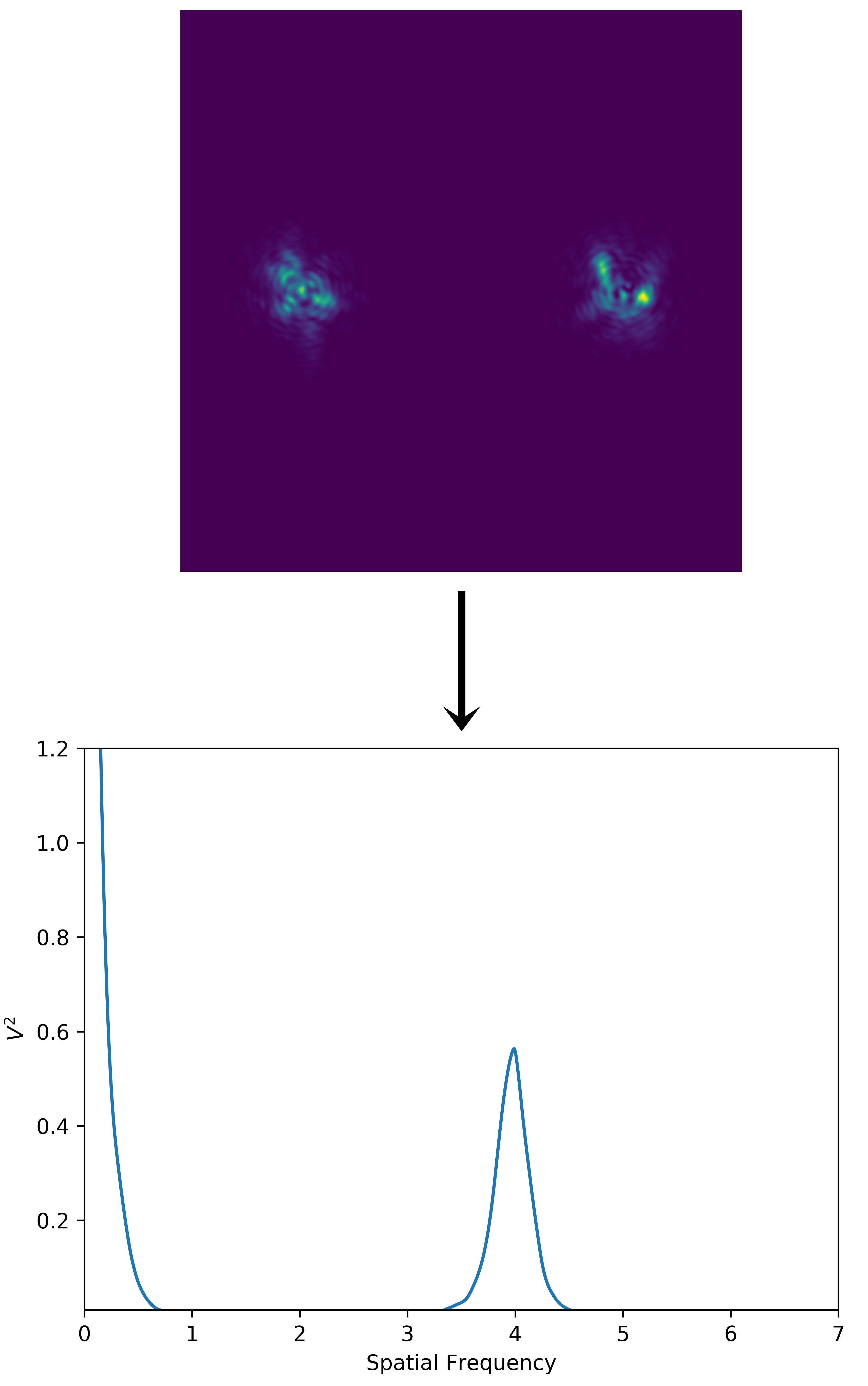}
    \caption{Two propagated pupils separated by four times their undiffracted diameters for a single exposure with the parameter configuration $\alpha$~=~3, $\beta$~=~3, $\delta$~=~0.1, $\gamma$~=~1, $T$~=~1. The upper image shows the instantaneous pupil intensity at the entrance to the beam combiner and the lower plot shows the resulting squared visibility of the above instantaneous pupil intensity.}
    \label{four_times_pupil_spacing}
\end{figure}

Throughout the rest of this paper, unless otherwise stated, the two diffracted pupils are separated by four times their undiffracted diameters as shown in Fig.~\ref{four_times_pupil_spacing}. This ensures both that the pupils do not overlap which would be nonphysical, and leaves a significant separation between the DC and interference term in the MTF. One exception here is for runs with $\beta$~$>$~4, in which case the pupils are separated just enough so they do not overlap, i.e.\ for runs with $\beta$~=~10 the pupils were spaced 10 times their undiffracted diameters. Unless otherwise stated the spatial frequency scale in all subsequent plots has been shifted such that the peak of the interference term is at a spatial frequency of zero.

\section{Results} \label{Results}

\subsection{Pupil shear} \label{Pupil_shear}

Low order atmospheric perturbations of tip and tilt are the dominant terms in an atmospherically perturbed wavefront and so are a natural first place to investigate. Tilt errors introduced by atmospheric perturbations can lead to significant beam shear in the beam combination laboratory due to the large propagation distances often experienced in long baseline optical interferometery. As the spatial frequency at which the interference term appears in the MTF depends on the spacing between the beams in the pupil plane, a naive model would assume that the spatial frequency of the interference term in the MTF would shift linearly with any change in pupil spacing and lead to the side lobes of the interference term also shifting and contributing to baseline crosstalk. 

\begin{figure}
	\includegraphics[width=\columnwidth]{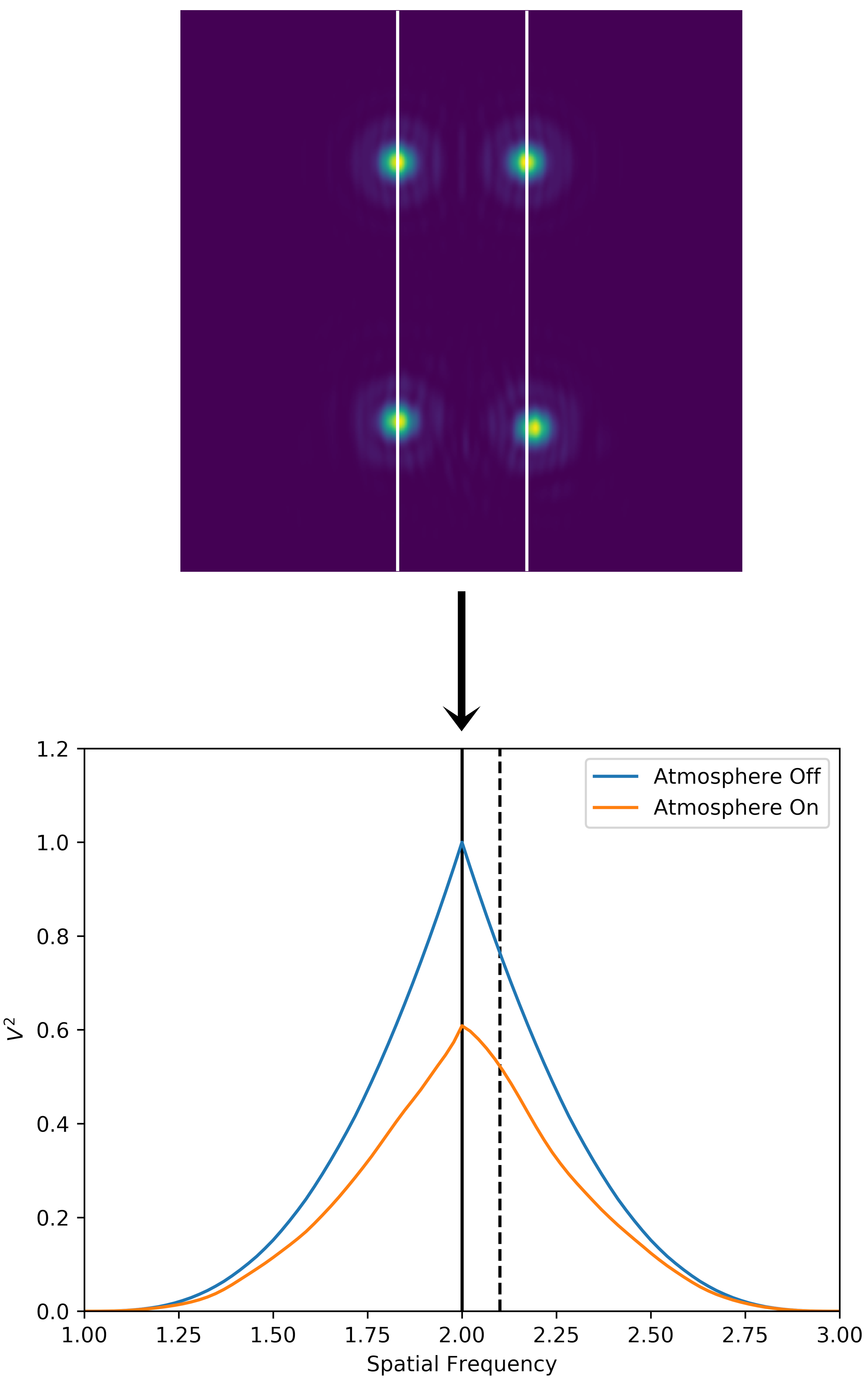}
    \caption{A demonstration that pupil shear due to propagation does not cause a shift in spatial frequency of the interference term. Top: the pupil plane after diffraction for a system with a $\delta$~=~0.5 and $\gamma$~=~1. The upper pair of pupils are the case of no atmospheric perturbations and are still separated by two times their undiffracted pupil diameters after propagation. The lower pair of pupils are from a separate run and super imposed and offset vertically with the same values of $\delta$ and $\gamma$ but with atmospheric perturbations at the level of $\alpha$~=~1.5. Due to tip-tilt induced shear errors these pupils are separated by 2.1 times their undiffracted pupil diameters. Bottom: the resulting Visibility for the two runs with the atmosphere off case giving a visibility of unity at a spatial frequency of 2 (solid vertical line). The atmosphere on case shows significant deviation from the ideal curve due to atmospheric perturbations but still peaks at a spatial frequency of 2. The dashed vertical line is at the spatial frequency the atmosphere on peak would be expected to be at based on the pupil separation alone.}
    \label{Pupil_shear_comparison}
\end{figure}

Our simulation shows that this is not the case however. In the ideal case of no cropping aperture ($\beta~\approx~\infty$) and equal path lengths ($\gamma$~=~1) we observe pupil shear altering the pupil spacing but the spatial frequency of the interference term does not shift. An example of this is given in Fig.~\ref{Pupil_shear_comparison} where a system with $\delta$~=~0.5 and $\gamma$~=~1 is compared with no atmospheric perturbations ($\alpha~=~0$) and atmospheric perturbations introduced at a level of $\alpha$~=~1.5 (atmosphere on). Note here the pupils are only separated by two times their undiffracted diameters. In the case of no atmospheric perturbations the propagated pupils remain spaced at two times their undiffracted diameters. The resulting visibility curve shows a visibility of unity at a spatial frequency of two. In the presence of atmospheric perturbations the pupil shear in the example shown changes the beam separation to 2.1 times the undiffracted beam diameter. The naive model of pupil shear would then expect the peak of the interference term to appear at a spatial frequency of 2.1 (dashed vertical line in the lower plot of Fig.~\ref{Pupil_shear_comparison}), however this is not the case and the peak still appears at a spatial frequency of two. 

For the example in Fig.~\ref{Pupil_shear_comparison} the simulation was reconfigured to remove Zernike modes with Noll indices between 3 and 15 from the atmospherically perturbed wavefront to isolate the effects of tip and tilt errors from higher order atmospheric perturbations. 

This discrepancy between the variation in the beam shear with propagation distance and the constancy of the spatial frequency of the fringe peak is a consequence of a more general result, which is that, under certain assumptions, the pattern resulting from the interference of light originating from a given set of telescope pupils is independent of the propagation distance. This more general result can be derived by considering the complex amplitude distribution $B(x,y)$ at the input to the beam combiner as the sum of a set of complex amplitude functions
\begin{equation}
    B(x,y)=\sum_j A_j(x,y,\delta_j)*\delta(x-x_j,y-y_j),
\end{equation}
where $A_j(x,y,\delta_j)$ is the
complex pupil amplitude function of the pupil from telescope $j$ after demagnification by a beam compressor and propagation by a dimensionless distance $\delta_j$ from the telescope, this is convolved with $\delta(x,y)$ which is the Dirac delta function and $x_j$ and $y_j$ are lateral offsets between the telescope pupils fixed by the design of the beam combiner. The complex amplitude $b(f_x,f_y)$ seen in the image plane is then given by a scaled version of the Fourier transform of the pupil function
\begin{equation}
    b(f_x,f_y)\propto \sum_j a_j(f_x,f_y,\delta_j)\exp(-2\pi i [x_j f_x + y_j f_y]),
\end{equation}
where $a_j(f_x,f_y,\delta_j)$ is the Fourier transform of $A_j(x,y,\delta_j)$. Equation~\eqref{propagation_equation} shows that propagation from the telescope to the beam combiner results in the multiplication of $a_j$ by a complex phase factor dependent on $\delta_j$. In the particular case of the propagation distance from all the telescopes being the same so that $\delta_j~=~\delta$, we get

\begin{multline}
    b(f_x,f_y)\propto \exp((-i\pi \delta)(f_{x}^{2} + f_{y}^{2}))\sum_j a_j(f_x,f_y,0) \\ \exp(-2\pi i [x_j f_x + y_j f_y]),
\end{multline}

which is independent of propagation distance except for a phase factor. The observed intensity pattern depends only on the modulus and not the phase of the complex amplitude in the image plane, so the interference pattern is independent of propagation distance, providing that the propagation distance is the same for all beams and there are no intervening apertures which block some of the propagating light.

The observation that the spatial frequency of the peak of the interference pattern does not shift even when the centre-to-centre separation of the pupils varies due to propagation indicates the need to consider both the amplitude and the phase of the pupil-plane pattern when deriving the autocorrelation of the pupil: shifts in the the pupil amplitude distribution caused by propagation effects must be accompanied by exactly compensating changes in the pupil phase distribution.

This result would imply that no crosstalk generated due to beam propagation, however as stated above this result is invalid if the propagation distance of the two pupils is not the same ($\gamma \neq $1) or for finite sized optics, ($\beta < \infty$). The remainder of this paper will explore these more general cases where propagation does affect crosstalk.

\subsection{Crosstalk dependence on \texorpdfstring{$\delta$}{delta} \& \texorpdfstring{$\beta$}{beta}} \label{beta_delta_discussion}

The effects of the propagation parameter $\delta$ and the aperture ratio parameter $\beta$ are presented simultaneously here as it is only when both $\beta$ is finite and $\delta>0$ that we see an effect on crosstalk. 

For a non-zero value of $\delta$, diffraction leads to an increase in the size of the beam. A finite value for $\beta$ leads to a cropping aperture acting as a form of spatial filter, leading to what \cite{2001MNRAS.327..217H} refers to as `Fresnel spatial filtering'. As the beam propagates, perturbations in the wavefront caused by atmospheric seeing on scales smaller than the diameter of the initial pupil will diffract out of the pupil faster than the pupil itself diffracts. When the beam is subsequently cropped, the higher spatial frequency perturbations are disproportionately removed, leaving the pupil partially spatially filtered. 

\begin{figure}
\begin{center}
	\includegraphics[width=0.75\columnwidth]{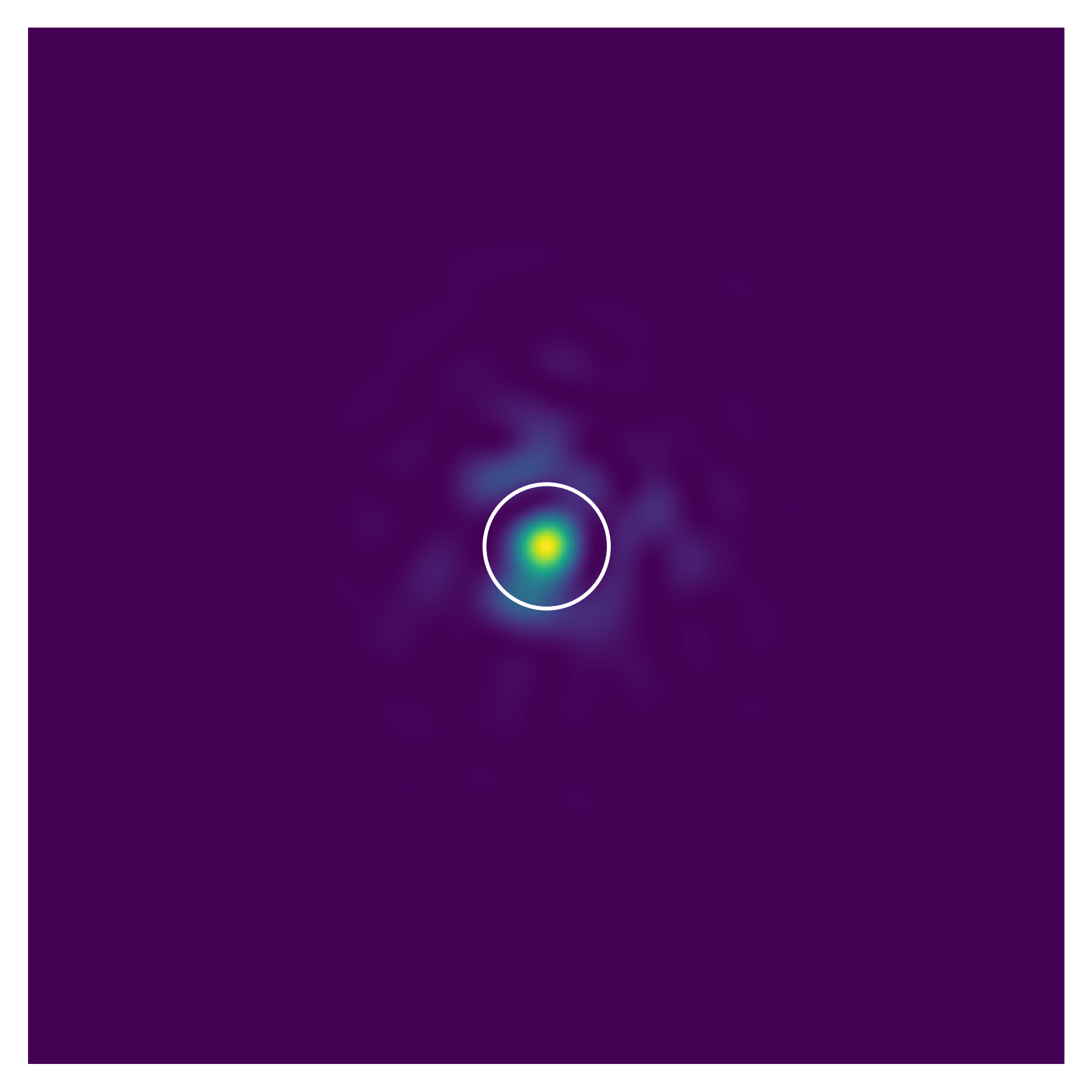}
    \caption{A pupil subject to atmospheric seeing ($\alpha$~=~4, $T$~=~1) after propagation by $\delta$~=~0.5, before being cropped by a cropping aperture of $\beta$~=~1 represented by the white ring. The higher spatial frequency amplitude perturbations arising from atmospheric seeing are disproportionately removed by the cropping aperture, leaving the pupil partially spatially filtered.}
    \label{Fresnel_filtering_pupil_example}
\end{center}
\end{figure}

This effect is demonstrated in Fig.~\ref{Fresnel_filtering_pupil_example} which shows a pupil subjected to atmospheric seeing ($\alpha$~=~4, $T$~=~1) after being propagated by $\delta$~=~0.5, before being cropped by a cropping aperture of $\beta$~=~1 (the size of the undiffracted pupil) represented by the white ring. The starlight which is kept after cropping (light within the white ring) is much more similar to what is expected from diffraction alone without atmospheric perturbations (top pupils in Fig.~\ref{Pupil_shear_comparison}) than the light which is cropped, i.e. the light retained contains fewer high spatial frequency perturbations arising from atmospheric seeing and so the beam is partially spatially filtered. 

\begin{figure}
	\includegraphics[width=\columnwidth]{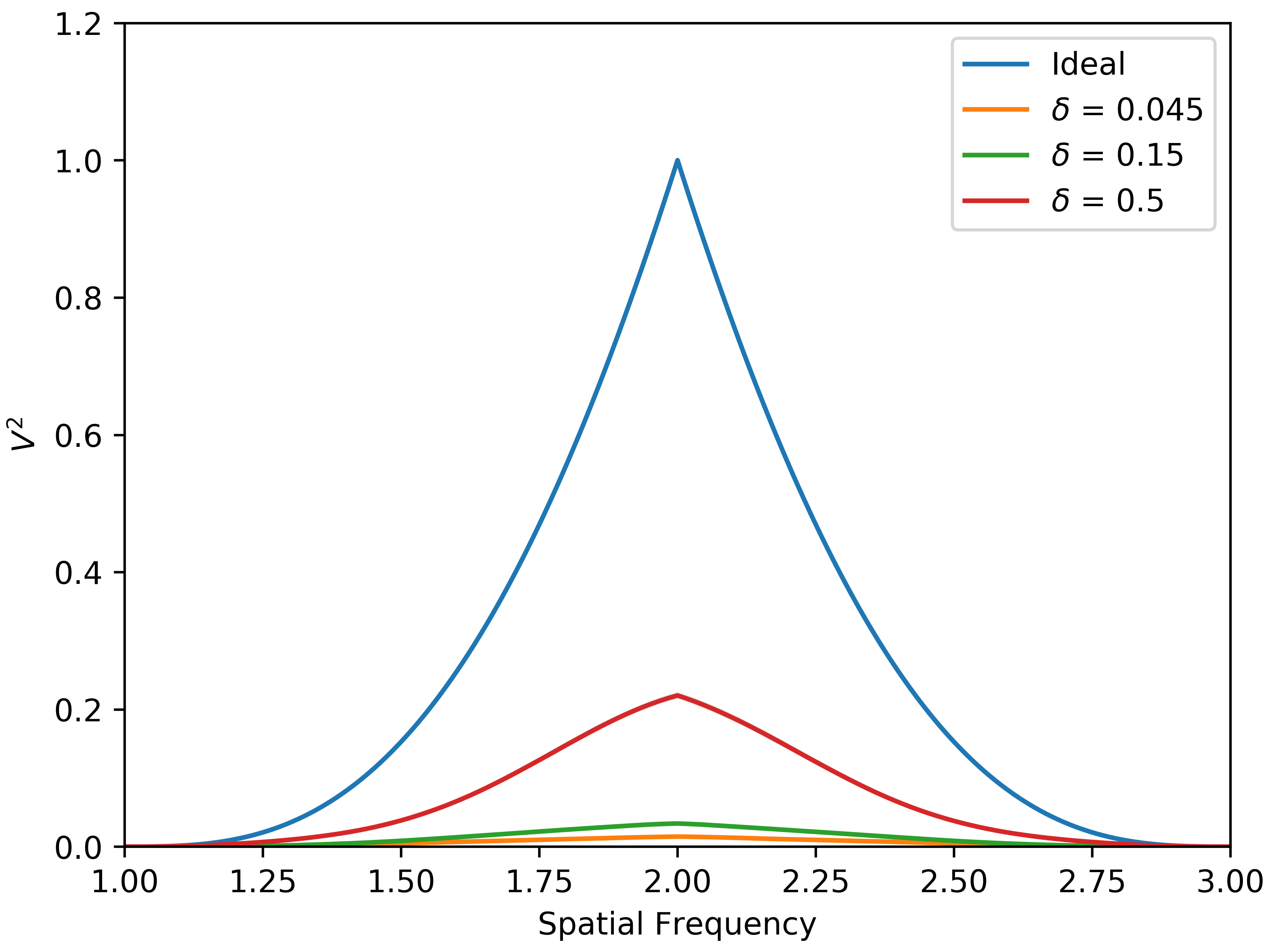}
    \caption{Visibility of the interference term for three values of $\delta$ showing an increase in visibility for increasing $\delta$ due to Fresnel spatial filtering. The other parameters were held constant at $\alpha$~=~9, $\beta$~=~1, $\gamma$~=~1, $T$~=~1. The spatial frequency axis has not been re-scaled and the peak of the DC term is at a spatial frequency of zero.}
    \label{delta_Fresnel_filtering}
\end{figure}

The further the beam propagates the greater the filtering. This effect is evident in our simulation as Fig.~\ref{delta_Fresnel_filtering} shows, the visibility of the interference term increases for a larger propagation distance given a fixed value of $\beta$.

\begin{figure}
	\includegraphics[width=\columnwidth]{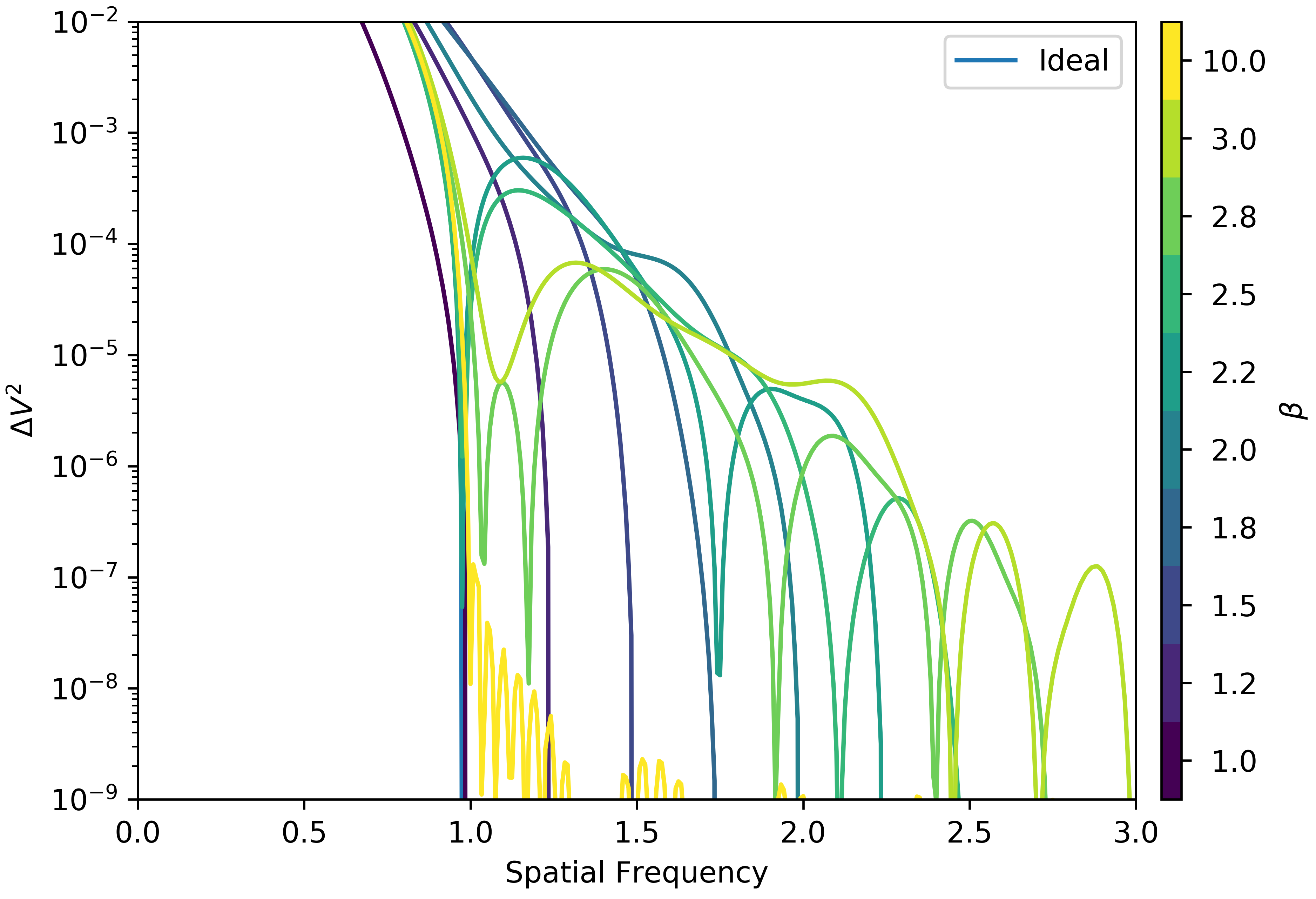}
    \caption{Crosstalk as a function of $\beta$ (cropping aperture size) for fixed parameter values of $\delta$~=~0.5, $\gamma$~=~1 when no atmospheric perturbations are induced. Here the crosstalk arises purely due to free space propagation. Note the tick marks indicate the values of $\beta$ plotted which follow the perceptually uniform colour map linearly with the exception of the $\beta$~=~10 run.}
    \label{atm_off_various_beta_for_one_config}
\end{figure}

Returning to the impact on crosstalk, we first consider the case of no atmospheric perturbations giving perfect plane wave initial pupils. As shown in Fig.~\ref{atm_off_various_beta_for_one_config}, we find non-zero MTF values above a spatial frequency of 1 for various values of $\beta$ (with the other relevant parameters fixed at $\delta$~=~0.5 and $\gamma$~=~1). These non-zero values mean that there can be crosstalk for aperture spacings such as that illustrated in Fig.~\ref{Auto_corr_sliding_mtf}, where the classical crosstalk between adjacent baselines is zero. In the following, we refer to any power beyond a spatial frequency of 1 as crosstalk, and give the resulting visibility value this crosstalk corresponds to the label $\Delta V^{2}$. Whether or not there is actual crosstalk in any given scenario depends on the spatial frequencies corresponding to different baselines.

\begin{figure}
	\includegraphics[width=\columnwidth]{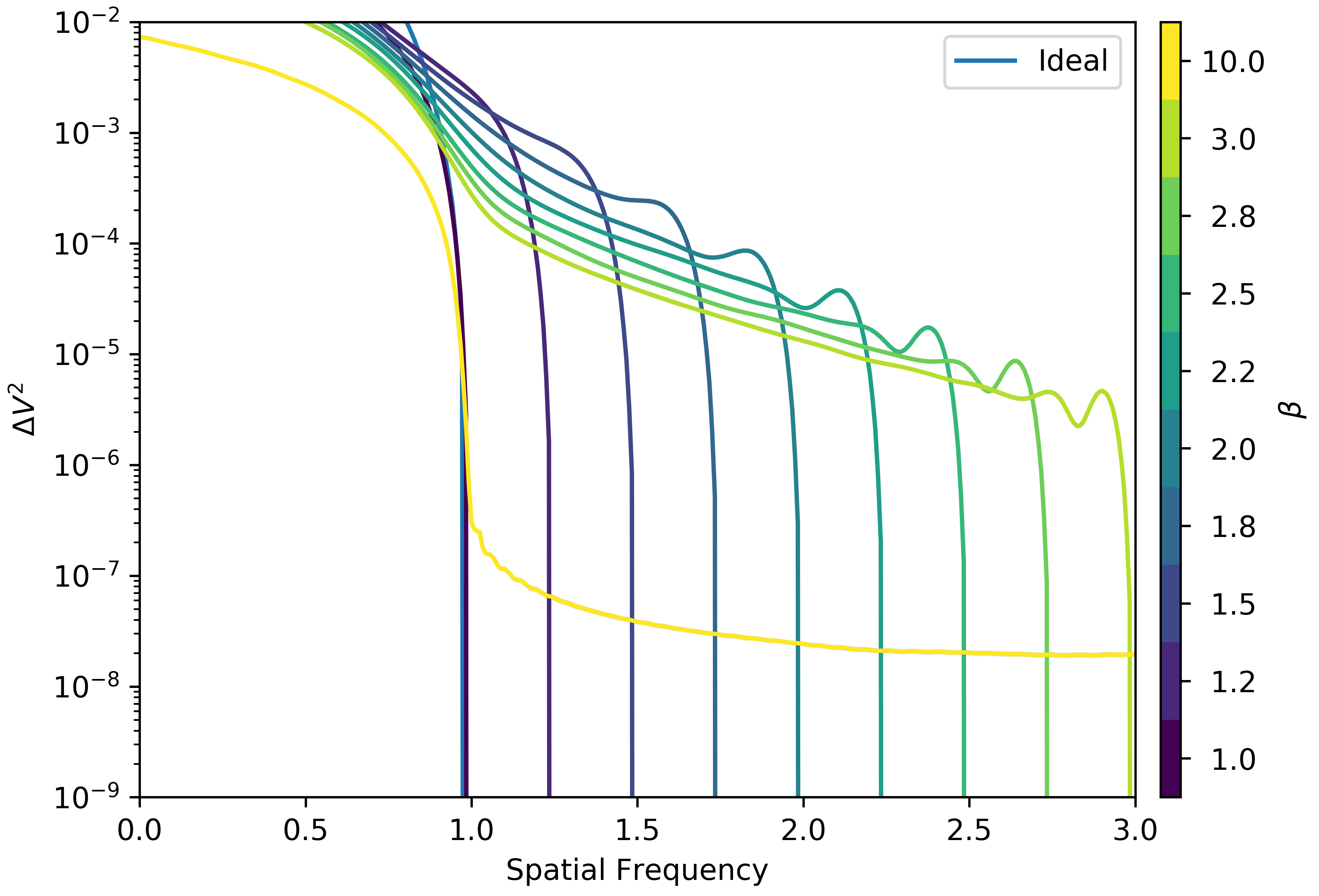}
    \caption{Crosstalk with atmospheric seeing effects as a function of $\beta$ with the other parameters held at $\alpha$~=~9, $\delta$~=~0.5, $\gamma$~=~1, $T$~=~0. Here the increasing extent of crosstalk for increasing values of $\beta$ is clear, with a sharp cut off at the maximum spatial frequency of crosstalk for a given value of $\beta$. Note the tick marks indicate the values of $\beta$ plotted which follow the perceptually uniform colour map linearly with the exception of the $\beta$~=~10 run.}
    \label{various_beta_for_one_config}
\end{figure}

Fig.~\ref{various_beta_for_one_config} shows that including finite amounts of atmospheric perturbations generally results in an increase in the level of crosstalk for a given value of $\beta$. There is a clear cut off in the maximum spatial frequency of the crosstalk for any given $\beta$. This effect can be understood by considering, as discussed in Section~\ref{Model}, that the MTF is the modulus of the autocorrelation of the pupil function. The area of the pupil expands due to diffraction however, for a value of $\beta$~=~1 as the pupils are cropped to their original diameters after propagation we have a similar situation as in Fig.~\ref{Auto_corr_sliding_mtf} where the MTF must be zero at spatial frequencies where there is no overlap of the pupils. As $\beta$ increases, the maximum pupil diameters and so the highest spatial frequency at which crosstalk can exist also increases. As the runs presented in Fig.~\ref{various_beta_for_one_config} are for a large value of $\delta$~=~0.5, the diffracted pupil is always larger than the cropping aperture and so it is the cropping aperture which limits the maximum spatial frequency resulting in the sharp cut off.

In addition to crosstalk at higher spatial frequencies as $\beta$ increases, the amplitude of the crosstalk decreases, tending towards near zero crosstalk for a large value of $\beta$. As $\beta$ tends towards infinity (an uncropped pupil) the system tends towards the case discussed in Section~\ref{Pupil_shear}, where the phase information aids in removing any crosstalk. This is evidenced by the $\beta$~=~10 run in Fig.~\ref{various_beta_for_one_config}.

\begin{figure}
	\includegraphics[width=\columnwidth]{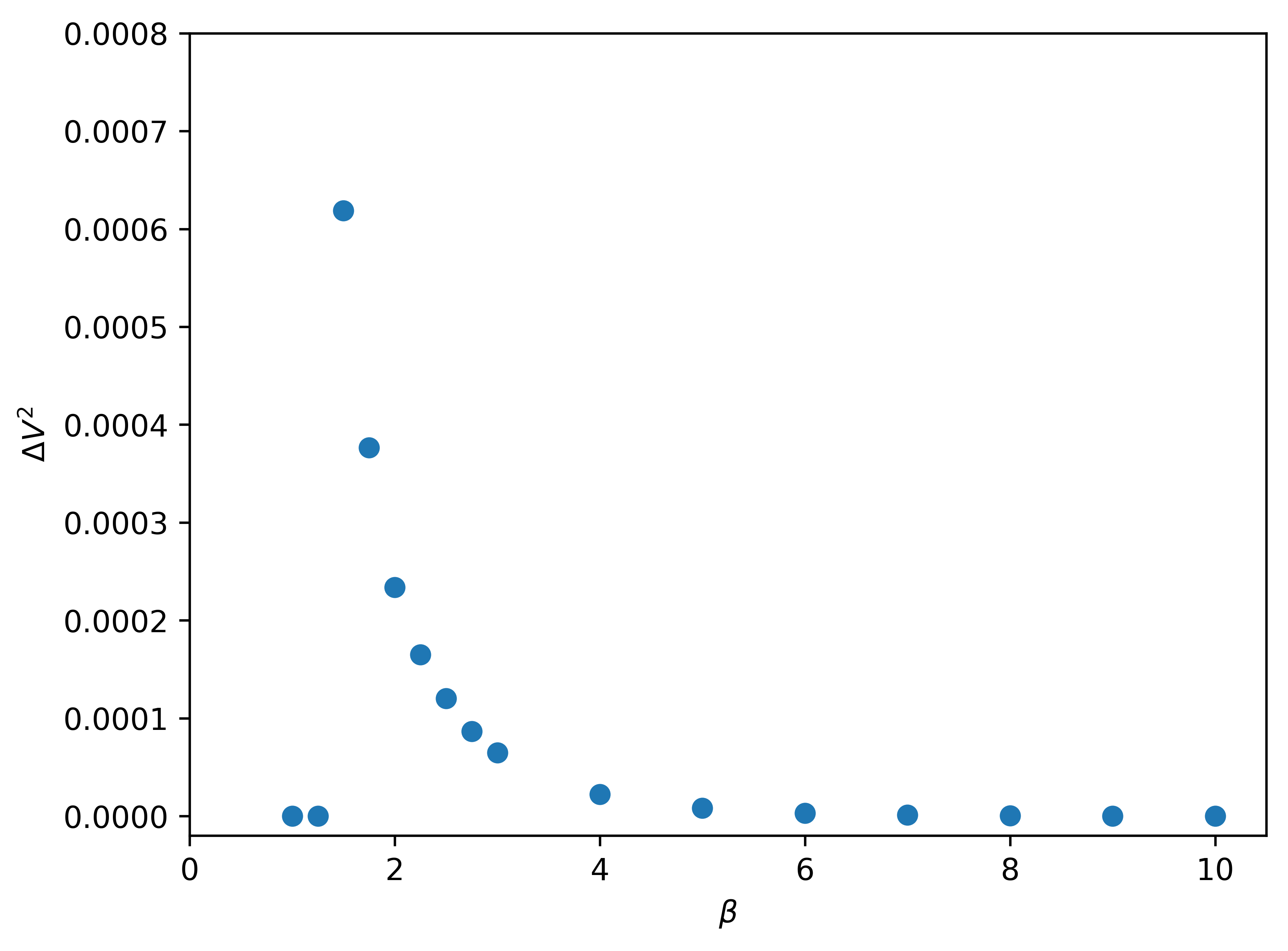}
    \caption{Crosstalk as a function of $\beta$ (cropping aperture size) at a spatial frequency of 1.3 with the other parameters held at $\alpha$~=~9, $\delta$~=~0.5, $\gamma$~=~1, $T$~=~0. Crosstalk is seen to be zero until the value of $\beta$ reaches 2 where it then peaks before decaying as $\beta$ increases further.}
    \label{various_beta_for_one_config_plotted_at_spatial_freq_1_3}
\end{figure}

These two effects lead to an interesting dependence of crosstalk at a fixed spatial frequency as a function of $\beta$ (cropping aperture size). As Fig.~\ref{various_beta_for_one_config_plotted_at_spatial_freq_1_3} shows for a fixed spatial frequency the level of crosstalk is zero until the cropped pupil is large enough to retain information at that spatial frequency (as per the discussion above in this section), at which point the crosstalk is at its maximum value. As $\beta$ continues to increase the amplitude of the crosstalk decays due to the system approaching the case of an uncropped pupil. In Fig.~\ref{various_beta_for_one_config_plotted_at_spatial_freq_1_3} the amplitude of the crosstalk is as high as $\Delta V^{2}$~=~\num{6.5e-4} for a value of $\beta$~=~1.5.

From these results we can conclude that both a cropping aperture equal to the diameter of the original beam ($\beta$~=~1) and an infinitely large cropping aperture induce no crosstalk. Neither of these cases may be practical, for example $\beta$~=~1 has the effect of reducing the throughput of an interferometer as fig.~4 of \citet{2001MNRAS.327..217H} shows, which may reduce the limiting magnitude of the array, and infinitely large optics are not practical. However the results do show that it is possible to avoid crosstalk originating from diffraction. When designing an image plane beam combiner the clear aperture of the optics along the beam train should be considered and the pupils spaced appropriately such that the maximum spatial frequency passed by the clear apertures does not overlap with the spatial frequency of any other baseline. 

This result should be considered alongside what values of $\beta$ and pupil spacings are physically possible. For example if the minimum pupil separation is twice the undiffracted diameter then the maximum physical value of $\beta$ is two as beyond this the cropping apertures would overlap. The maximum physical value of $\beta$ scales with beam separation, for example if the pupils were separated by three times their undiffracted diameters the maximum physical value of $\beta$ would be three and so on.

\begin{figure}
	\includegraphics[width=\columnwidth]{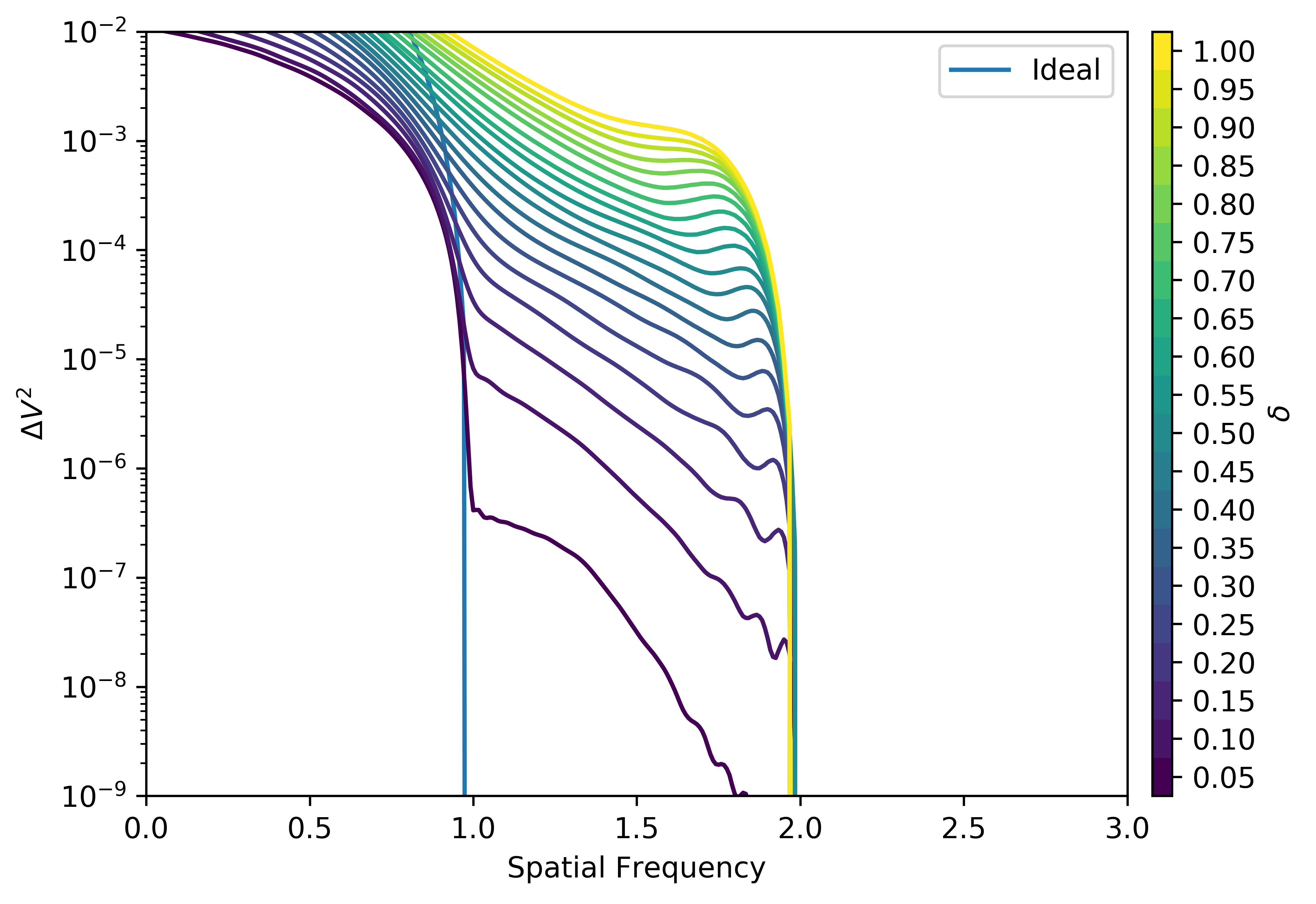}
    \caption{Crosstalk for a range of $\delta$ values with the other parameters held at $\alpha$~=~9, $\beta$~=~2, $\gamma$~=~1, $T$~=~1. Increasing $\delta$ increases the crosstalk amplitude at all spatial frequencies which display diffraction crosstalk.}
    \label{delta_Alpha_9_Beta_2_Gamma_1_Tip_tilt_1}
\end{figure}

Exploring the effects of varying $\delta$ we find that, while $\beta$ may define the profile of the crosstalk, $\delta$ defines the amplitude of it as demonstrated in Fig.~\ref{delta_Alpha_9_Beta_2_Gamma_1_Tip_tilt_1} by the clear increase in crosstalk amplitude with increasing $\delta$. We find that larger values of $\delta$ result in greater amplitudes of crosstalk, with the exception of runs for which $\beta$~=~1 which do not display any crosstalk.

\begin{figure}
	\includegraphics[width=\columnwidth]{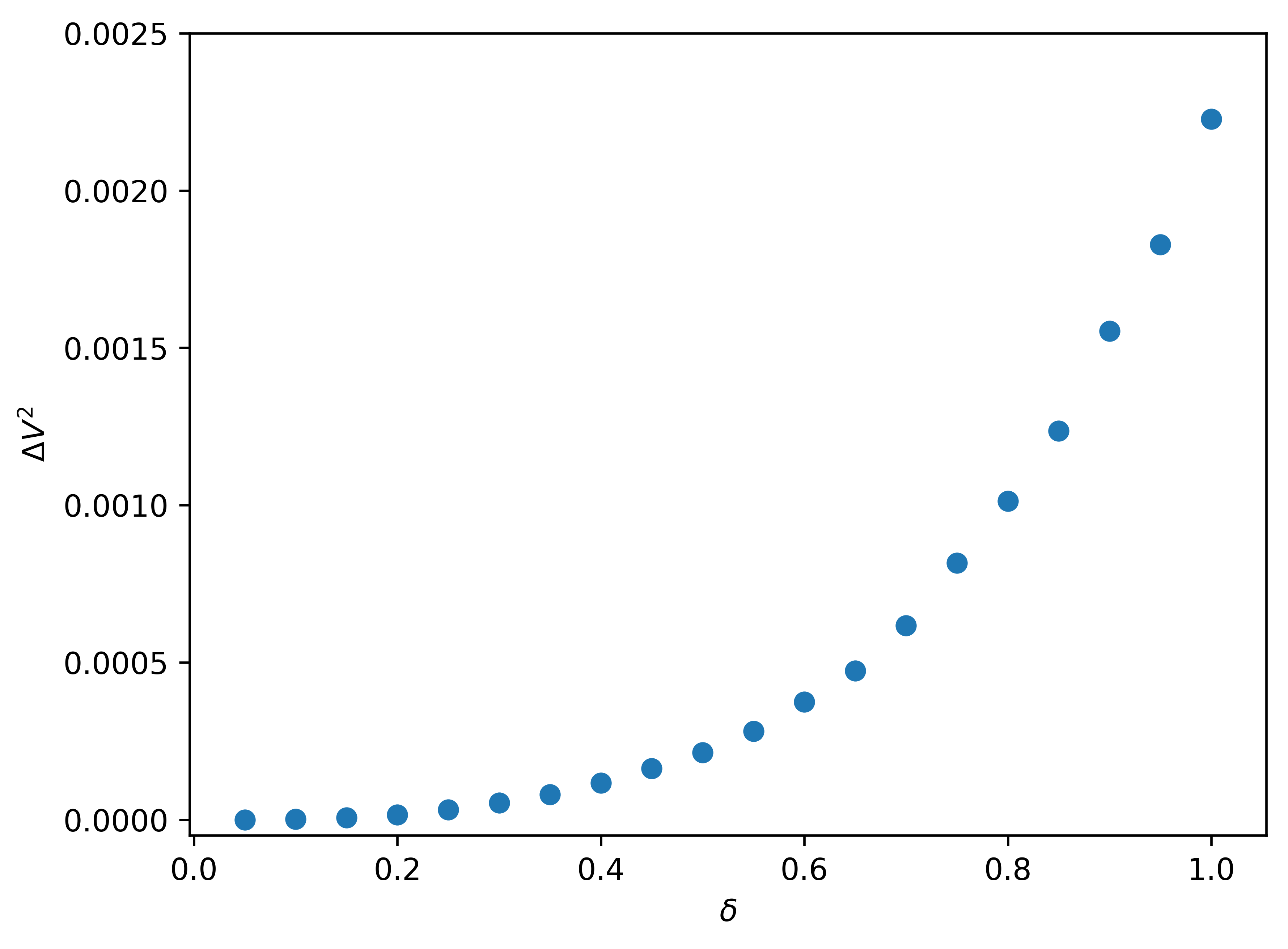}
    \caption{Crosstalk amplitude as a function of $\delta$ (impact of diffraction) at a spatial frequency of 1.3 with the other parameters held at $\alpha$~=~9, $\beta$~=~2, $\gamma$~=~1, $T$~=~1.}
    \label{as_func_delta_fixed_spatial_freq}
\end{figure}

In Fig.~\ref{as_func_delta_fixed_spatial_freq} we explore the crosstalk amplitude as a function of $\delta$ (impact of diffraction) for a fixed set of other parameters ($\alpha$~=~9, $\beta$~=~2, $\gamma$~=~1, $T$~=~1) at a spatial frequency of 1.3, showing the crosstalk amplitude monotonically increases as a function of $\delta$. It should be noted that the simulations presented here where $\delta$>0.5 the pupil is sampled more coarsely at 60 pixels across the diameter as opposed to 120 pixels used elsewhere. 

These results show that it is important to consider the diffraction parameter $\delta$ when estimating crosstalk, with larger values of crosstalk arising for observations at longer wavelengths and larger propagation distances. 

\subsection{Crosstalk dependence on \texorpdfstring{$\alpha$}{alpha}}

\begin{figure}
	\includegraphics[width=\columnwidth]{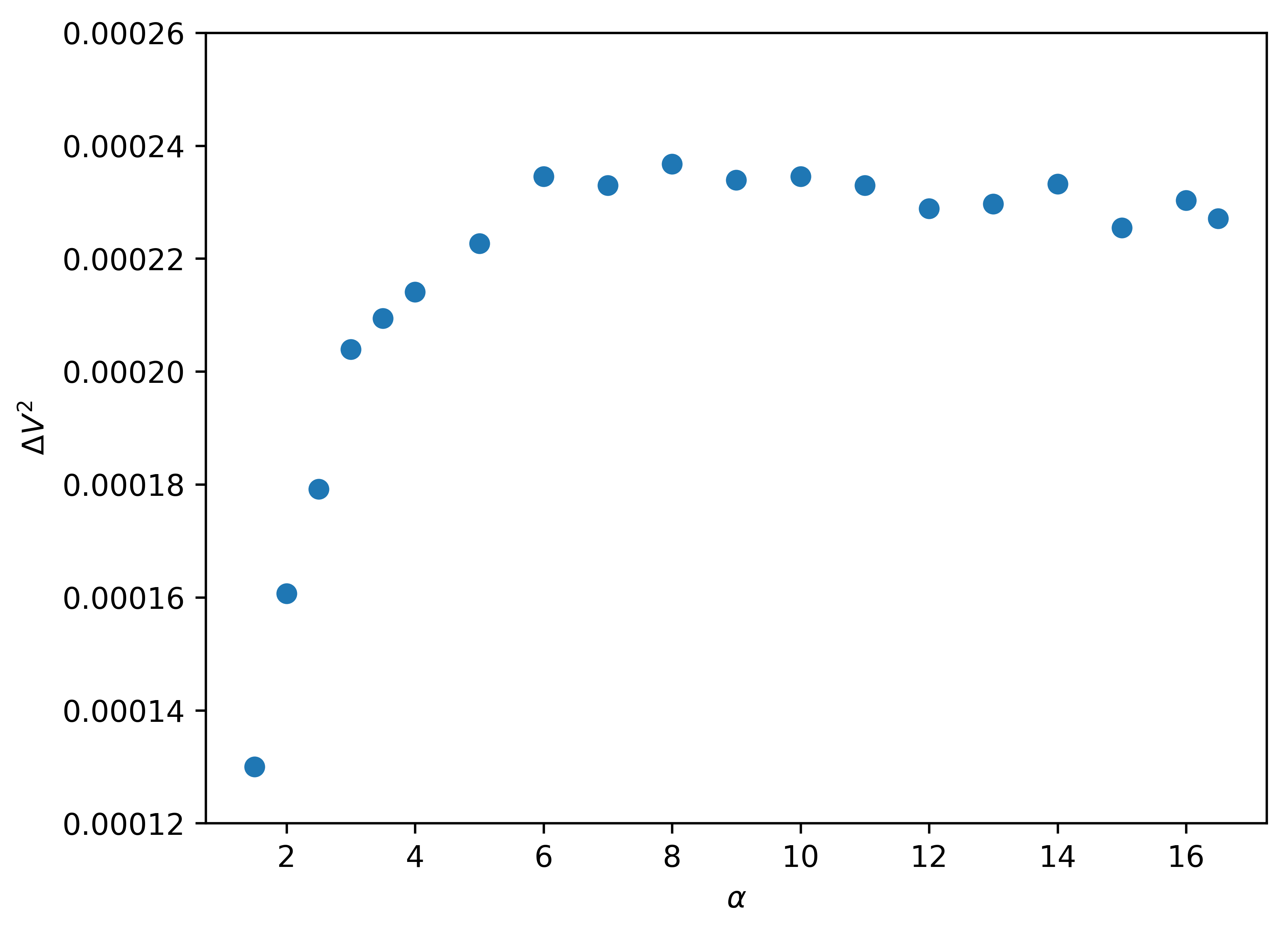}
    \caption{Crosstalk as a function of $\alpha$ (level of seeing) at a spatial frequency of 3.5 for $\beta$~=~2, $\delta$~=~0.5, $\gamma$~=~1, $T$~=~0. The amplitude of crosstalk is seen to approximately increase linearly up to a value of $\alpha$~=~4 before flattening out with no significant increase in crosstalk seen between $\alpha$~=~8 and $\alpha$~=~16.5.}
    \label{Alpha_one_config_freq_3.5}
\end{figure}

\begin{figure}
	\includegraphics[width=\columnwidth]{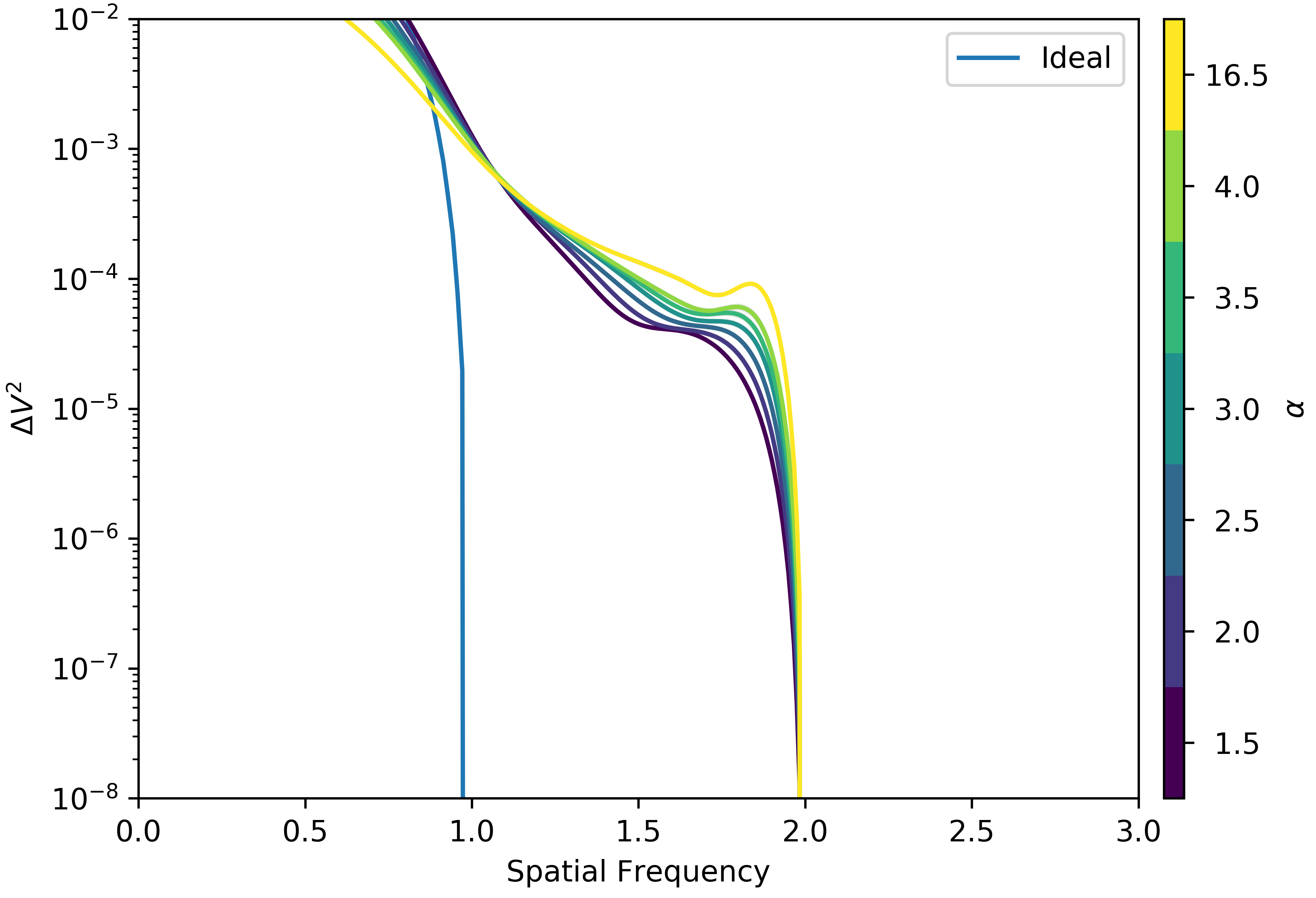}
    \caption{Crosstalk for various values of $\alpha$ (level of seeing) with the other parameters held at $\beta$~=~2, $\delta$~=~0.5, $\gamma$~=~1, $T$~=~0. Here the increase in amplitude of crosstalk for increasing values of $\alpha$ at all spatial frequencies up to a value of $\alpha$~=~4 is evident. Note the tick marks indicate the values of $\alpha$ plotted which follow the perceptually uniform colour map linearly with the exception of the $\alpha$~=~16.5 run.}
    \label{Alpha_one_config}
\end{figure}

For a given configuration of parameters, a larger value of the seeing parameter $\alpha$ results in a greater amplitude of crosstalk. Fig.~\ref{Alpha_one_config_freq_3.5} shows the amplitude of crosstalk at a fixed spatial frequency as a function of $\alpha$ which increases up to a value of $\alpha$~=~4. Above $\alpha$~=~4 the amplitude of crosstalk begins to level off with no significant increase between $\alpha$~=~8 and $\alpha$~=~16.5. A select number of runs for varying levels of $\alpha$ are shown in Fig.~\ref{Alpha_one_config} showing that this trend is true at all spatial frequencies which exhibit crosstalk. 

These results suggest that in the regime below a value of $\alpha$~=~4 any reduction in the value of $\alpha$ will significantly reduce the crosstalk. However above a value of $\alpha$~=~8 there will be no significant reduction in the amplitude of the crosstalk if $\alpha$ is reduced. This result is of relevance if schemes to calibrate the level of crosstalk are contemplated, because they indicate the level of stability of the calibration in the presence of fluctuating seeing levels.

\subsection{Crosstalk dependence on \texorpdfstring{$T$}{T}}

\begin{figure}
	\includegraphics[width=\columnwidth]{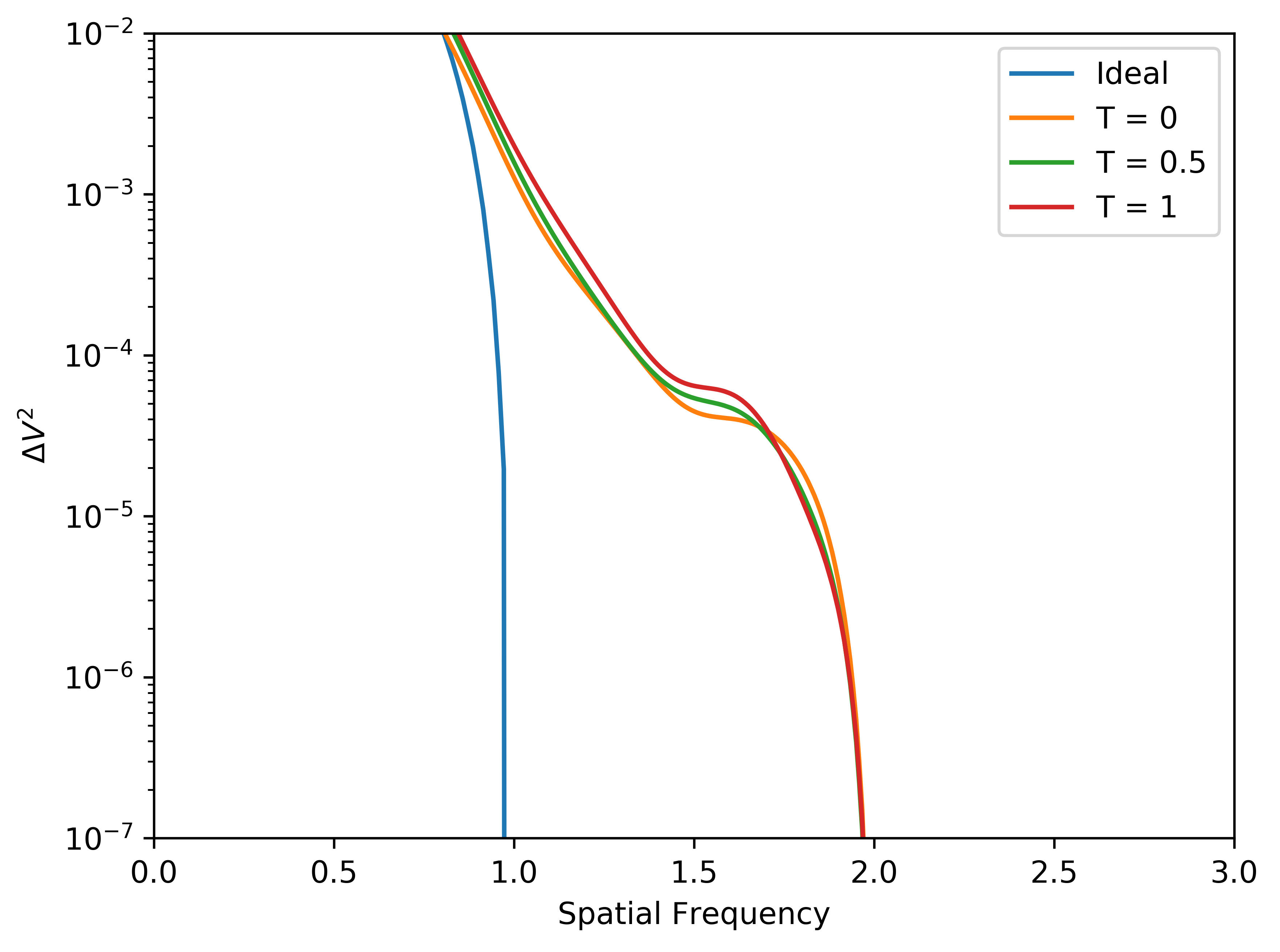}
    \caption{Varying levels of $T$ (tip-tilt correction) with the other parameters held at $\alpha$~=~1.5, $\beta$~=~2 $\delta$~=~0.5, $\gamma$~=~1. Here the variation in profile of crosstalk for differing levels of tip-tilt correction can be seen.}
    \label{Tip_tilt_alpha_1_5_beta_3_delta_0_5_gamma_1}
\end{figure}

\begin{figure}
	\includegraphics[width=\columnwidth]{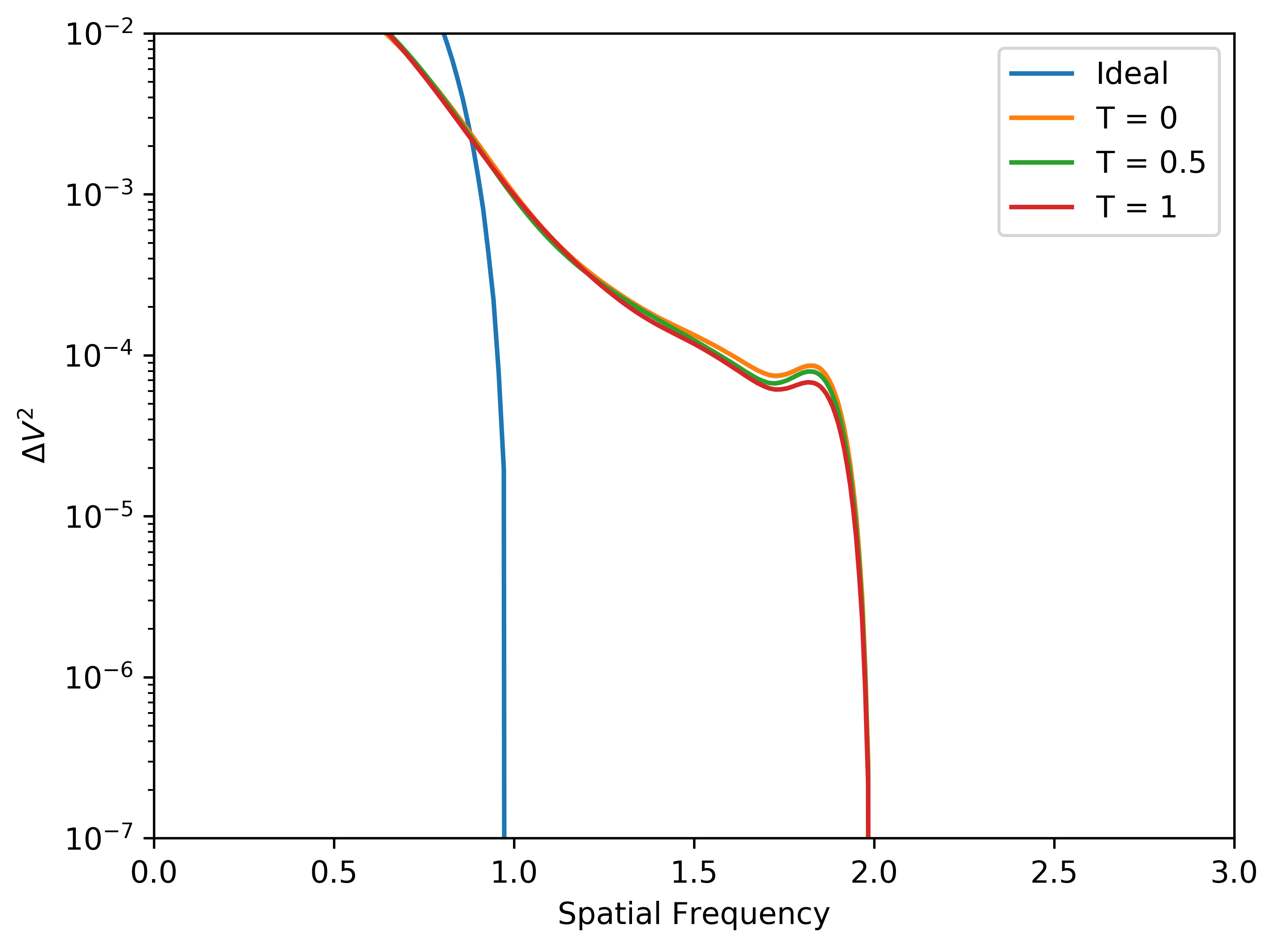}
    \caption{Varying levels of $T$ (tip-tilt correction) with the other parameters held at $\alpha$~=~9, $\beta$~=~2, $\delta$~=~0.5, $\gamma$~=~1. Here the minimal effect of tip-tilt correction on the observed crosstalk at larger values of $\alpha$ can be seen.}
    \label{Tip_tilt_alpha_9_beta_2_delta_0.5_gamma_1}
\end{figure}

Fig.~\ref{Tip_tilt_alpha_1_5_beta_3_delta_0_5_gamma_1} shows how changing the parameter $T$, encoding the level of tip-tilt corrections affects crosstalk. 
We see that tip-tilt corrections have a larger effect on the level of crosstalk for smaller values of the seeing parameter $\alpha$. Looking at a value of $\alpha$~=~1.5 in Fig.~\ref{Tip_tilt_alpha_1_5_beta_3_delta_0_5_gamma_1} we see the variation in the tip-tilt parameter $T$ for a given configuration does have a impact on the crosstalk profile whereas for $\alpha$~=~9 in Fig.~\ref{Tip_tilt_alpha_9_beta_2_delta_0.5_gamma_1}, with otherwise the same parameter configuration, the various levels of tip-tilt correction have significantly less of an impact.

This can be explained heuristically as follows: as atmospheric seeing worsens, then for any given level of seeing the relative power of the Zernike polynomials within the atmospheric phase distortions remain the same, with tip and tilt accounting for around 90\% of the total power. However, once the total level of the residual (non-tip-tilt) RMS wavefront error reaches about one radian, at $\alpha\sim 2$, then whether or not the tip-tilt fluctuations are removed becomes less important because the effect of additional aberrations is attenuated. A similar effect is seen in fig. 1 of \cite{1966JOSA...56.1372F} where the effect of tip-tilt correction diminishes for $D$/$r_0$~>~2.

\subsection{Crosstalk dependence on \texorpdfstring{$\gamma$}{gamma}}

From the definition of the propagation distance ratio parameter $\gamma$ given in Section~\ref{Parameter_space}, for a fixed value of $\delta$, a larger value of $\gamma$ corresponds to a greater difference in the propagation distance of the two beams. This results in a larger mismatch in the profile of the two pupils. The effect of this difference in profile of the two beams is that the two pupils are less coherent with each other and hence produce a weaker interference term at all spatial frequencies, including in the regime of crosstalk. 

\begin{figure}
	\includegraphics[width=\columnwidth]{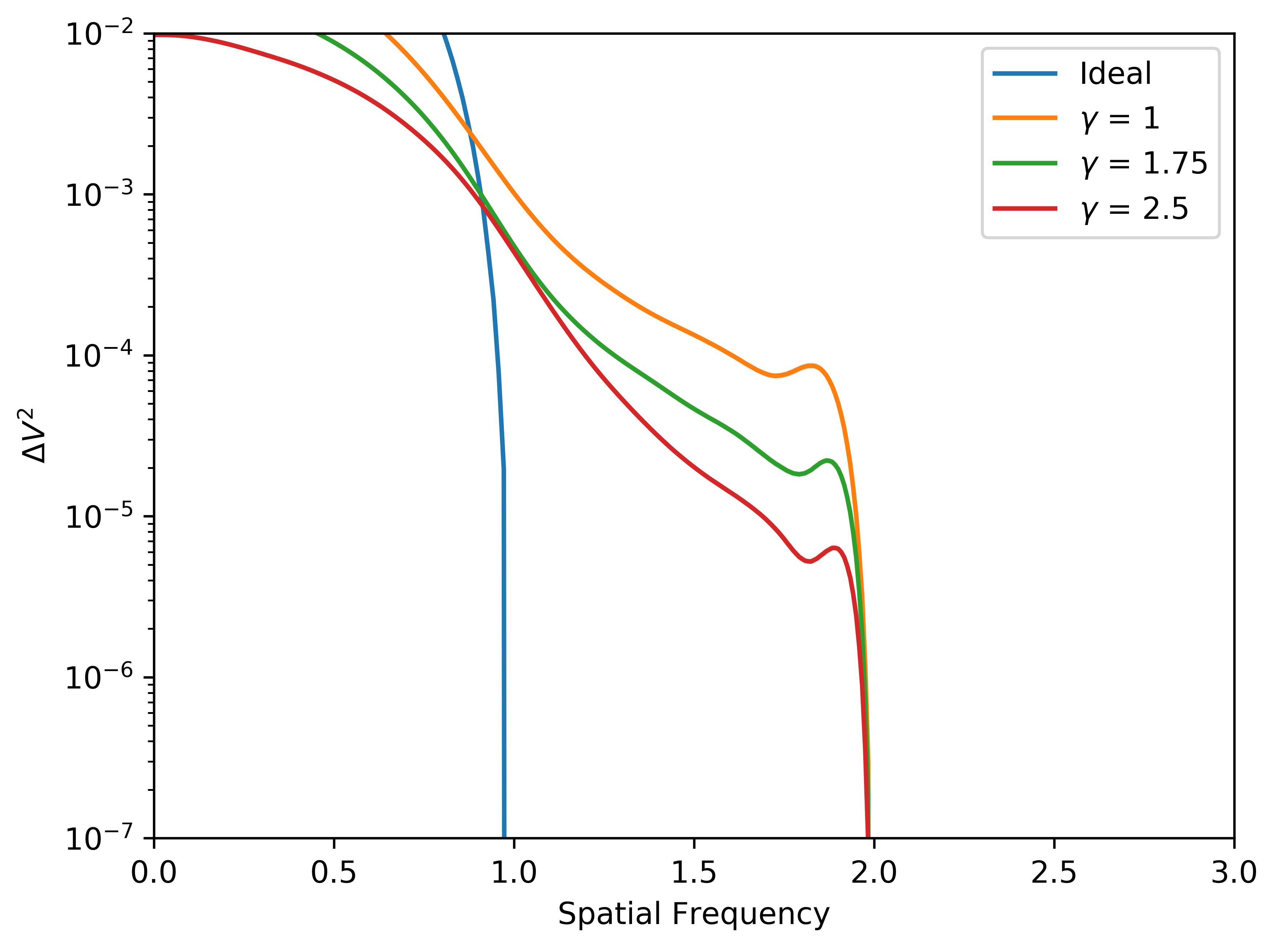}
    \caption{Various values of $\gamma$ (propagation path length mismatch) for one configuration with the other parameters held at $\alpha$~=~9, $\beta$~=~2, $\delta$~=~0.5, $T$~=~0. Here there is a decrease in crosstalk for a larger $\gamma$ due to a greater difference in the profile of the two beams at larger values of $\gamma$.}
    \label{gamma_alpha_9_beta_2_delta_0.5_T_0}
\end{figure}

This is seen in our simulation in Fig.~\ref{gamma_alpha_9_beta_2_delta_0.5_T_0} which shows that for a given configuration a larger value of $\gamma$ significantly decreases the level of crosstalk.

\begin{figure}
	\includegraphics[width=\columnwidth]{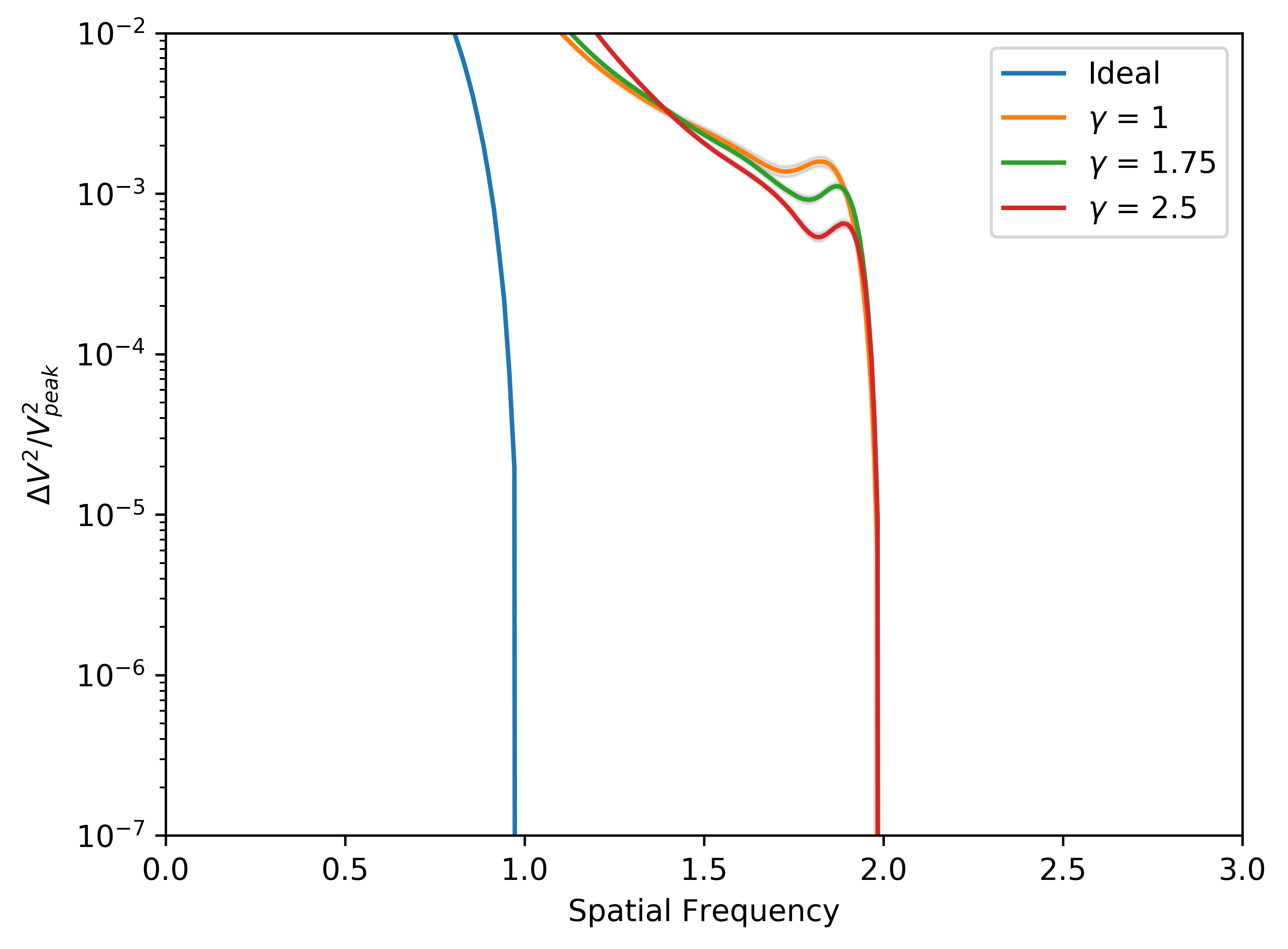}
    \caption{The visibility of the crosstalk relative to visibility at the peak of the interference term as a function of $\gamma$ (propagation path length mismatch) with the other parameters held at $\alpha$~=~9, $\beta$~=~2, $\delta$~=~0.5, $T$~=~0.}
    \label{gamma_rel_to_peak_alpha_9_beta_2_delta_0_5_T_0}
\end{figure}

In Fig.~\ref{gamma_rel_to_peak_alpha_9_beta_2_delta_0_5_T_0} the visibility of the crosstalk term is divided by the value of the visibility at the peak of the interference term to remove the effects of loss of coherence for greater values of $\gamma$. In this case the profiles of the three $\gamma$ values are significantly closer than in Fig.~\ref{gamma_alpha_9_beta_2_delta_0.5_T_0}. This demonstrates that the visibility of the crosstalk decreases by approximately the same amount as the interference term for a larger value of $\gamma$, suggesting the decrease in crosstalk with $\gamma$ is due to a reduction in coherence. Hence while the crosstalk term may decrease, its amplitude relative to the interference term remains approximately the same. Thus if baselines at adjacent spatial frequencies have similar values of $\gamma$ then the effects of crosstalk on the measured visibilities is roughly independent of $\gamma$. 

\section{Discussion} \label{Discussion}

In this section we explore the implication of the results of this paper on real world examples of both a three and five telescope image plane combiner. 

\subsection{Three Telescope Beam Combiner} \label{Three_scope_combiner}

Here we apply the results of this paper to the FOURIER beam combiner at the MROI \citep{2020SPIE11446E..0VM}. FOURIER is a three beam J, H and K band image plane combiner which utilises a 2,4,6 beam spacing configuration (as in  Fig.~\ref{fft_mod_ideal_PSF_three_beams_2_4_6_seperation_combined}). FOURIER will cover the wavelength range of $\lambda$~=~\SI{1.1}{\micro\metre}--\SI{2.4}{\micro\metre} on baselines of B~=~\SI{7.8}{\metre}--\SI{374}{\metre}.

The data reduction pipeline for FOURIER is still under development however one method for extracting the visibility is to apply equation~\eqref{Vis_equation} to an ensemble of interferograms and extract the visibility at the spatial frequency of the peak of the interference term as predicted by the pupil plane spacing for each baseline. We are confident in this method as the result in Section~\ref{Pupil_shear} has shown that the spatial frequency of the peak of the interference term in the MTF does not change even when atmospheric tilt induces pupil shear at the entrance of the beam combiner. 

From Section~\ref{Results} we can see that the worst case scenario would be for the largest values of $\delta$ and $\beta$, and an $\alpha$ value greater than eight. Tip-tilt correction and $\gamma$ were shown to be weak variables on the amplitude of crosstalk. Hence the worst case scenario will be for an observation on the longest baseline of \SI{347}{\metre}, 30$^\circ$ above the horizon, at a wavelength of $\lambda$~=~\SI{2.4}{\micro\metre}. 

The \SI{347}{\metre} baseline at the MROI is formed by two telescopes at the edge of two of the arms of the Y shaped array configuration, giving a propagation distance from one telescope to the centre of the array to be \SI{200.3}{\metre}, the distance from the centre of the array to the delay line entrance is an additional \SI{50}{\metre} of propagation. The delay line imparts a delay of \SI{300.5}{\metre} to one of the beams of starlight to compensate for the geometric delay given by $\tau~=~B\cos(\phi)$, where $B$ is the baseline and $\phi$ the angle of the target above the horizon. An additional \SI{10}{\metre} of propagation is added for the delay line exit to beam compressor distance. This then is the path length over which the beam propagates while being \SI{95}{\milli\metre} in diameter, giving a contribution of $\delta$~=~0.189. The beam is then compressed to \SI{13}{\milli\metre} in diameter and propagated a further \SI{14.19}{\metre} from the beam compressors to the FOURIER beam combiner, adding an additional $\delta$~=~0.257. This gives the final value of $\delta$ to be $\delta$~=~0.446.

Given that the delay line must introduce a \SI{300.5}{\metre} delay to one of the beams to compensate for the geometric delay, with the path length otherwise being identical for the two beams, this observation has a value of $\gamma$~=~1.29. This value of $\gamma$ is calculated by a slightly different definition to that given in equation~\eqref{e:gamma} as it is the difference in how affected by propagation the two beams are, rather than the ratio of their absolute path length we are concerned with. This distinction is often not necessary and is only needed here as the beam is compressed in diameter for the final \SI{14.19}{\metre} of propagation. By taking $\gamma$ to be the ratio of the two $\delta$ values for the two beams rather than the absolute path length we arrive at a value of $\gamma$~=~1.29. 

Assuming a seeing of 1 arcsecond at a wavelength of \SI{0.5}{\micro\metre} at zenith and using the scaling relations from Section~\ref{Parameter_space} gives $\alpha$~=~3.14. While this is a moderately low value of $\alpha$ and we would expect greater crosstalk for larger values of $\alpha$ (in this case observations at shorter wavelengths) by comparing Figs.~\ref{as_func_delta_fixed_spatial_freq} and \ref{Alpha_one_config_freq_3.5} we see that crosstalk is a much stronger function of $\delta$ than $\alpha$ and so the worst case scenario is still for the maximum $\delta$ with a smaller $\alpha$. Finally we assume in this run that tip-tilt aberrations are perfectly corrected, $T$~=~1.

\begin{figure}
	\includegraphics[width=\columnwidth]{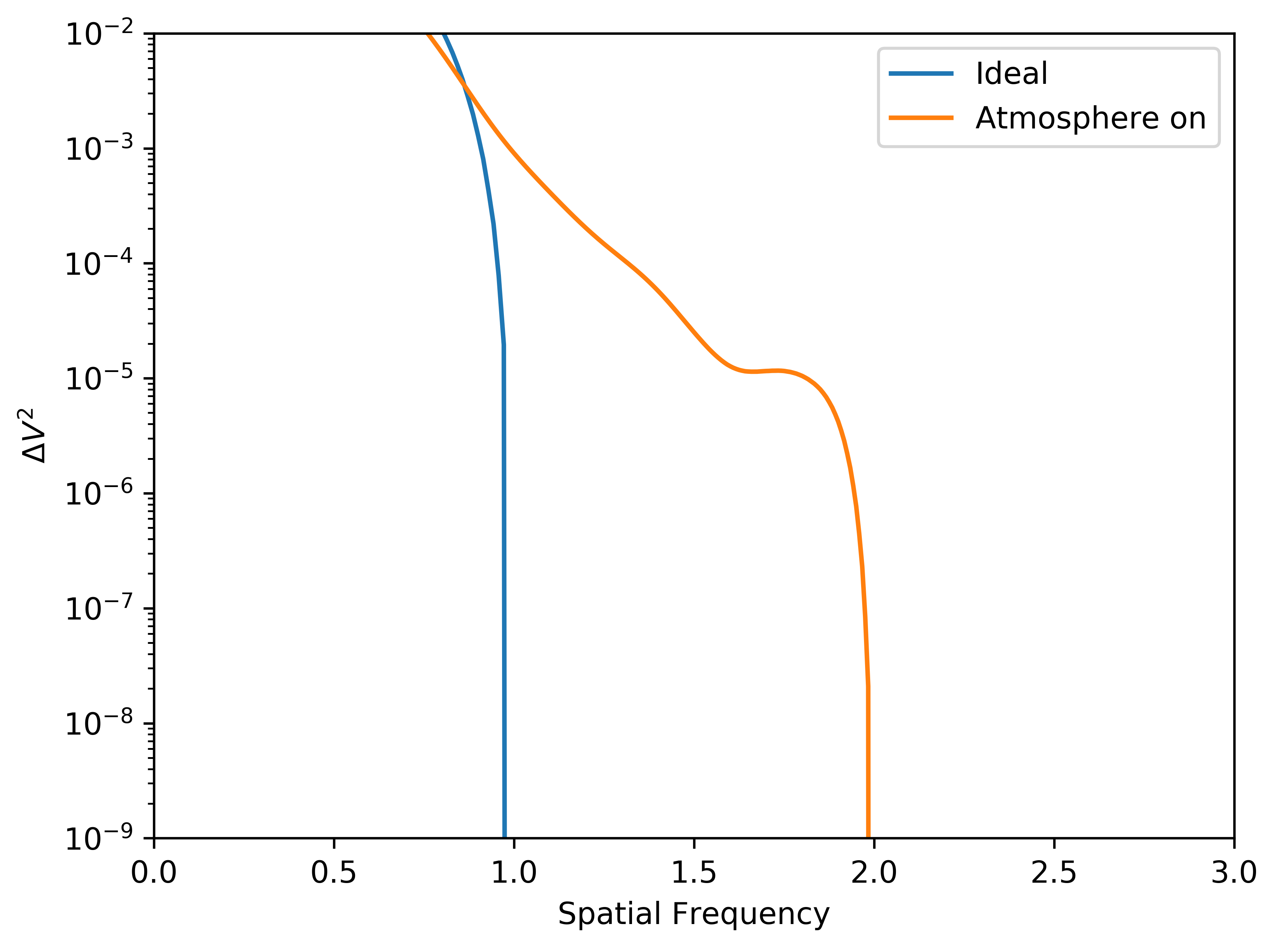}
    \caption{Crosstalk for the expected worst case scenario observations at the MROI ($\alpha$~=~3.14, $\beta$~=~2, $\delta$~=~0.446, $\gamma$~=~1.29, $T$~=~1) compared to the ideal case. There is significant crosstalk beyond the classical cut off at a spatial frequency of one however this goes to zero at a spatial frequency of two due to the cut off for $\beta$~=~2.}
    \label{Alpha_3_14_Beta_2_Delta_0_446_Gamma_1_29_Tip_tilt_1}
\end{figure}

The 2,4,6 beam spacing configuration used by FOURIER is the same as used in Fig.~\ref{fft_mod_ideal_PSF_three_beams_2_4_6_seperation_combined}. The smallest beam spacing utilised in FOURIER is two times the undiffracted aperture and so the maximum value of $\beta$ that can be used is two. As discussed in Section~\ref{beta_delta_discussion}, when $\beta$~=~2 the highest spatial frequency diffraction-induced crosstalk can reach would be just below one spatial frequency unit higher than what would be expected if no crosstalk were present, i.e.\ diffraction induced crosstalk from the lowest spatial frequency baseline would reach zero just before a spatial frequency of four in Fig.~\ref{fft_mod_ideal_PSF_three_beams_2_4_6_seperation_combined}. This is demonstrated in Fig.~\ref{Alpha_3_14_Beta_2_Delta_0_446_Gamma_1_29_Tip_tilt_1} which shows the crosstalk does reach zero at a spatial frequency just below one spatial frequency unit higher than the ideal case. Considering only the effects discussed so far this would mean the diffraction induced crosstalk would be zero for FOURIER due to the well spaced pupils.

\begin{figure}
	\includegraphics[width=\columnwidth]{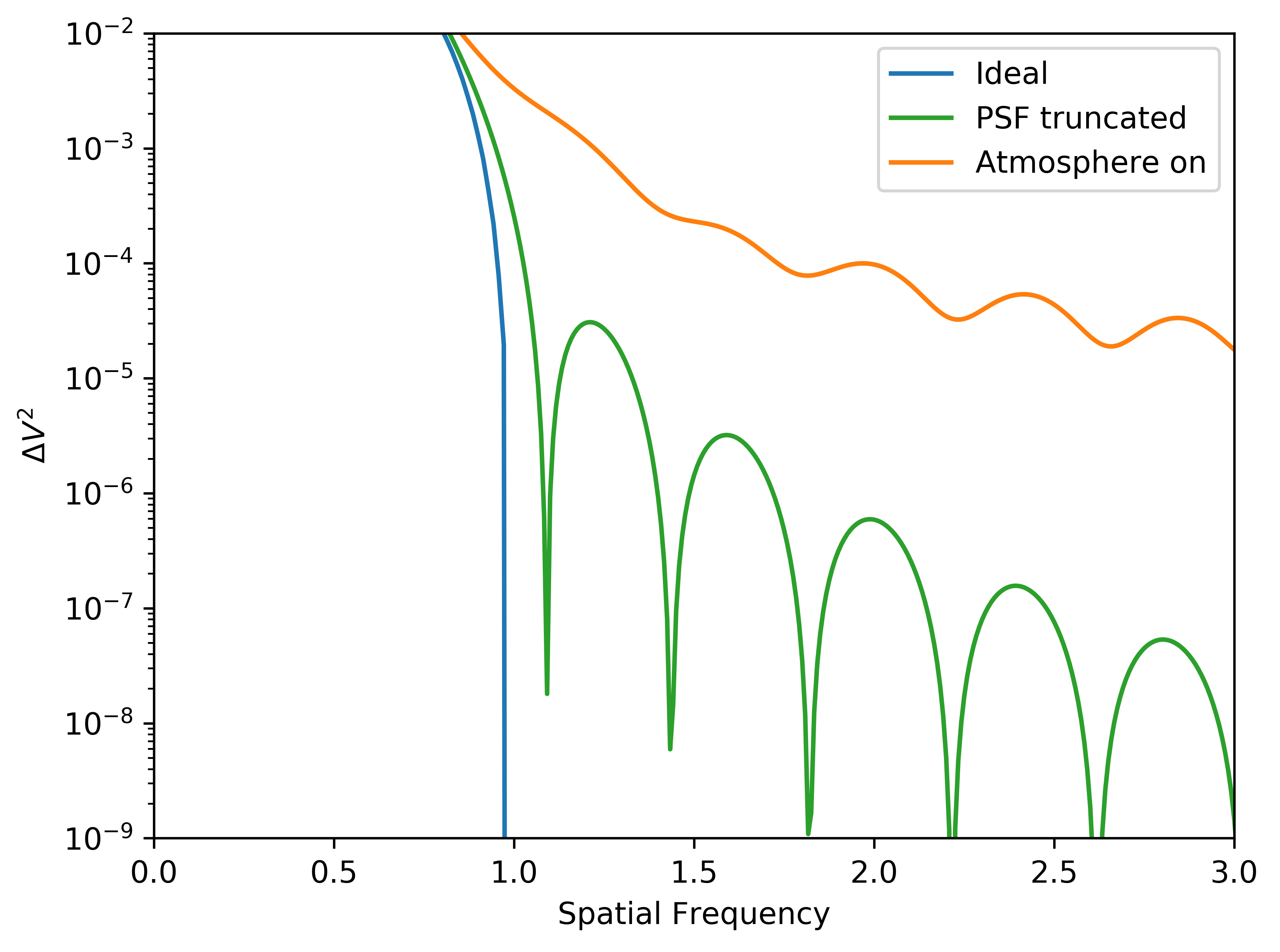}
    \caption{Crosstalk when the PSF is truncated at the first null of the Airy disk as is the case when pinhole spatial filtering is implemented. The effects of truncation with no propagation or atmospheric effects and for the expected worst case scenario of FOURIER are shown by the `PSF truncated' and `Atmosphere on' lines, respectively.}
    \label{PSF_truncation_Alpha_3.14_Beta_2_Delta_0.446_Gamma_1.29_Tip_tilt_1_pixels_per_ro_38.22}
\end{figure}

One additional effect to consider is the use of spatial filtering by a pinhole spatial filter as will be implemented for FOURIER. Here the PSF is truncated typically at the location of the first null of the Airy disk. The effect of this is to multiply the PSF by a circular function of amplitude unity within the pinhole and zero outside. By the convolution theorem a multiplication in the image plane space results in a convolution of the Fourier transform of the two functions in the MTF space as the MTF is the modulus of the Fourier transform of the PSF.  Consider the 1D case where the circular function is represented by a rectangular function, in this case the ideal MTF is convolved with a sinc function with decaying side lobes out to infinite spatial frequency. To demonstrate the effects of this convolution the MTF of the ideal PSF (with no atmospheric or propagation effects) with and without truncating the PSF at the first null of the Airy disk is shown in Fig.~\ref{PSF_truncation_Alpha_3.14_Beta_2_Delta_0.446_Gamma_1.29_Tip_tilt_1_pixels_per_ro_38.22}. The truncation does indeed result in decaying side lobes at spatial frequencies well beyond the cut off frequency of the ideal case. 

Fig.~\ref{PSF_truncation_Alpha_3.14_Beta_2_Delta_0.446_Gamma_1.29_Tip_tilt_1_pixels_per_ro_38.22} also shows the effects of PSF truncation for the worst case scenario observation at the MROI in the `Atmosphere on' line. The crosstalk now extends well beyond the cut off when truncation is not considered but for an otherwise identical run to that shown in Fig.~\ref{Alpha_3_14_Beta_2_Delta_0_446_Gamma_1_29_Tip_tilt_1}. At the spatial frequency of the next baseline for FOURIER (two spatial frequency units higher than the previous baseline) the crosstalk is now $\Delta V^{2}$~=~\num{9.8e-5}. For the central baseline where there is an adjacent baseline both sides of it the situation is worsened as the crosstalk is symmetric and extends to both higher and lower spatial frequencies. The result of which is that the crosstalk for this baseline is doubled to $\Delta V^{2}$~=~\num{2.0e-4}.

\subsection{Five Telescope Beam Combiner}

The number of baselines and hence interference terms in the MTF goes as $N_b~=~(n^2 - n)/2$, where $N_b$ is the number of baselines and $n$ is the number of telescopes in the array. Hence the number of interference terms increases rapidly for higher order beam combiners, a five telescope beam combiner will produce 10 unique baselines, each of which must still be sampled at a unique spatial frequency. When considering the pupil spacing it is desirable to minimise the largest spatial frequency a baseline is sampled at, both to reduce the overall width of the linear non-redundantly spaced pupils, and to minimise the range of spatial frequencies which must be sampled. The higher the largest spatial frequency the larger the number of pixels needed in order to sample at least two pixels per fringe cycle (ideally four or above), and using more pixels both increases the size of detector needed and the effect of read noise as a greater number of pixels must be read out overall. 

\begin{figure}
	\includegraphics[width=\columnwidth]{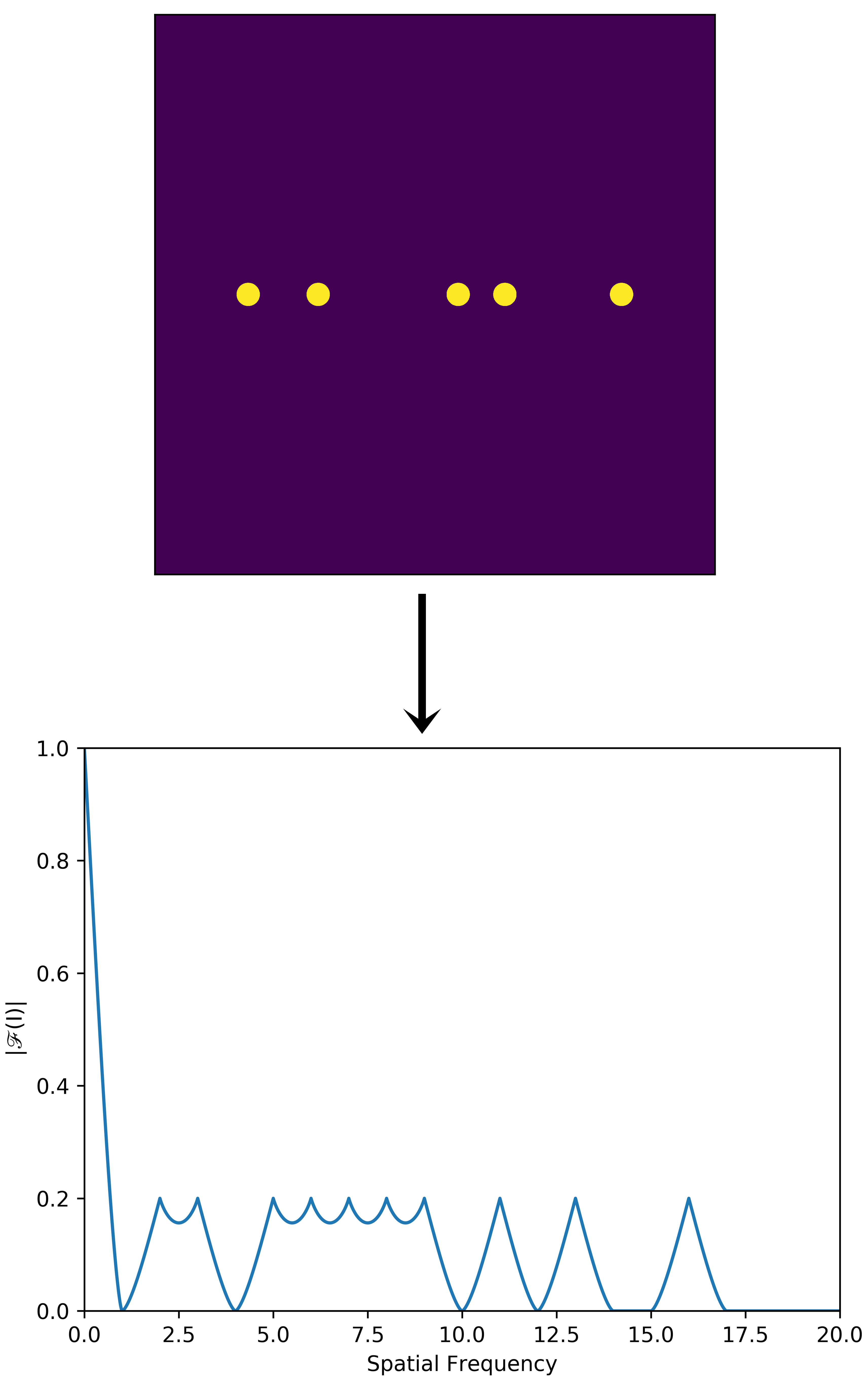}
    \caption{Top: five pupils spaced 3, 9, 11, 16 pupil diameters from the left most pupil. This configuration minimises the total width of the pupil spacing and hence the highest baseline spatial frequency. Bottom: the resulting MTF showing the 10 baselines generated from the five pupils all sampled at unique spatial frequencies.}
    \label{fft_mod_ideal_PSF_0_3_9_11_16}
\end{figure}

Fortunately the problem of minimising the total width of a non-redundantly spaced linear array has already been tackled and solutions are known as Golomb rulers \cite{sidon1932satz, babcock1953intermodulation}. We are interested in particular in optimal Golomb rulers which give the shortest total width for a non-redundant spacing pattern \citep{2001PASP..113...98G}. Optimal rulers for five telescope configurations are 11 pupil diameters in length, two of these rulers are given by placing pupils 1, 4, 9, 11 and 2, 7, 8, 11 pupil diameters away from the first pupil (which is placed at zero). Unfortunately the shortest spacing between a pupil pair for these configurations is only one pupil diameter restricting the maximum value of $\beta$ to $\beta$~=~1 which is often not practical and would place the lowest spatial frequency term at the null of the DC term. The next best solution is to take a optimal ruler for six telescopes and remove the pupil which forms the lowest spatial frequency term. For example a spacing configuration of 3, 9, 11, 16 pupil diameters away from the first pupil which is placed at zero. This gives the pupil plane configuration and MTF shown in Fig.~\ref{fft_mod_ideal_PSF_0_3_9_11_16} where all baselines are separated by at least one unit of spatial frequency from the adjacent baselines in the MTF which in the ideal case as described in Fig.~\ref{Auto_corr_sliding_mtf} would result in no baseline crosstalk. 

Using the parameter configuration laid out in Section~\ref{Three_scope_combiner} we see from Fig.~\ref{Alpha_3_14_Beta_2_Delta_0_446_Gamma_1_29_Tip_tilt_1} that one spatial frequency unit higher gives a crosstalk amplitude $\Delta V^{2}$~=~\num{9e-4} from the adjacent baseline. As for the three telescope combiner, for baselines where there is an adjacent baseline both sides of it the situation is worsened as the crosstalk is symmetric and extends to both higher and lower spatial frequencies. The result of which is that for three of the baselines in the configuration shown in Fig.~\ref{fft_mod_ideal_PSF_0_3_9_11_16} the crosstalk will be as high as $\Delta V^{2}$~=~\num{1.8e-3}.

If spatial filtering is also implemented in the five telescope case, Fig.~\ref{PSF_truncation_Alpha_3.14_Beta_2_Delta_0.446_Gamma_1.29_Tip_tilt_1_pixels_per_ro_38.22} shows that the amplitude of the crosstalk at one spatial frequency unit higher increases to $\Delta V^{2}$~=~\num{3.3e-3}. For baselines where there is an adjacent baseline on both sides in the MTF the crosstalk will be as high as $\Delta V^{2}$~=~\num{6.6e-3}.

As we have show, Fig.~\ref{Alpha_3_14_Beta_2_Delta_0_446_Gamma_1_29_Tip_tilt_1} and Fig.~\ref{PSF_truncation_Alpha_3.14_Beta_2_Delta_0.446_Gamma_1.29_Tip_tilt_1_pixels_per_ro_38.22} are significant results of this paper, highlighting that the additional crosstalk terms discussed in this paper can have a impact on real world observations. 

\section{Conclusions} \label{Conclusions}

In this work we investigated image plane crosstalk arising due to previously unexplored effects of atmospheric seeing, free space beam propagation, finite sized optics, unequal path lengths and tip-tilt correction. We find moderate levels of crosstalk ($\Delta V^{2}$~=~\num{6.6e-3}) due to these effects at spatial frequencies which were previously considered unaffected by crosstalk. Exploring the parameter space we find: 

(i) As the ratio of the diameter of the cropping aperture to the pupil diameter ($\beta$) tends towards infinity, pupil shear caused by tip-tilt atmospheric perturbation coupled with none-zero free space propagation distances do not cause the interference term to shift in spatial frequency.

(ii) The highest spatial frequency at which crosstalk is present (if spatial filtering is not implemented) is determined by the diameter of the cropping aperture, with larger cropping apertures enabling crosstalk at higher spatial frequencies.

(iii) A larger value of the propagation parameter $\delta$ results in a greater amplitude of crosstalk for a given system.

(iv) For a given system, the amplitude of crosstalk steadily increases as a function of the seeing parameter $\alpha$ up to a value of around $\alpha$~=~4, beyond which the amplitude of crosstalk levels out with no significant difference between 8 $\leq~\alpha~\leq$ 16.5. 

(v) Tip-tilt adaptive optics corrections have only a limited impact on the amplitude of the crosstalk explored here, though its effects are greater at lower values of $\alpha$.

(vi) There is a decrease in the amplitude of the crosstalk for larger values of the propagation ratio parameter $\gamma$, due to a reduction in coherence between the two pupils as they propagate by more different distances giving a larger mismatch in the pupil profiles. 

(vii) Crosstalk rises rapidly as interference terms in the MTF are brought closer together. It is relatively easy to arrange the pupil plane for three beam combiners to avoid diffraction crosstalk however, it must be considered more carefully when designing higher order (five telescopes or more) beam combiners. We recommend the use of `crosstalk resilient modes' where the interference terms are spaced at distances in the MTF greater than the maximum spatial frequency the crosstalk can reach for the value of $\beta$ being used. 

(viii) Care must be taken when implementing spatial filtering via a pinhole which, when coupled with diffraction crosstalk, causes a significant amount of crosstalk at spatial frequencies far from the interference term itself.  

\section*{Acknowledgements}

We thank the reviewer and editor for their timely and insightful comments, which have improved the manuscript.

DJM acknowledges a PhD studentship from the UK Science and Technology Facilities Council.

\section*{Data Availability}

The data underlying this article are available on Github, at \href{https://github.com/dmortimer101/Crosstalk_optical_interferometers}{https://github.com/dmortimer101/Crosstalk\_optical\_interferometers}.



\typeout{}
\bibliographystyle{mnras}
\bibliography{references} 



\bsp	
\label{lastpage}
 \end{document}